# Entering the Elite, But Remaining Outside Its Core? The Temporal Structure of Gender Inequality in the Global Citation Elite

**Marek Kwiek**
Center for Public Policy Studies, Adam Mickiewicz University, Poznan, Poland
kwiekm@amu.edu.pl, ORCID: orcid.org/0000-0001-7953-1063

## Abstract

This study analyzed the gender dynamics of the global citation elite using the annual Top 2% Stanford/Elsevier lists for 2019–2024. While previous analyses have discussed women's representation (inclusion), I discuss what happens once male and female scientists are within the elite (position and persistence). The database used included 1.22 million person-year observations (N = 1,216,089) and 465,707 unique scientists and scholars from more than 150 countries. I studied entry, exit, and permanent membership in the elite. The share of women in the top 2% increased from 18.35% in 2019 to 20.85% in 2024. However, women were more strongly represented among first-observed entrants than among continuing members, and their share decreased from 21.58% among single-year members to 17.69% among scientists present in all six annual lists. Women were located closer to the lower boundary of the elite in terms of citation index deciles (the lowest decile included 26.88% of women). Men, in contrast, were closer to its upper boundary (the highest decile included only 15.23% of women). Logistic regression models estimated a lower probability of women's membership in the durable core of the elite (odds ratio estimate [OR] = 0.886). Consequently, increasing the inclusion of women in the elite does not mean their equal position within or equal persistence of membership in it. I draw conclusions about temporal gender dynamics within the global citation elite and gender inequalities in science more generally.

# 1. Introduction

Gender inequalities in science are well documented, and women are underrepresented in many highly visible and prestigious segments of the global science system. The differences between men and women are shown across academic disciplines, countries, and career stages, and are documented with respect to authorship patterns, citations, self-citations, research funding, and access to scientific recognition (Huang et al., 2020; Larivière et al., 2013; Meho, 2021; Ross et al., 2022; Sánchez-Jiménez et al., 2024). They are also visible among the scientists located at the top of the global citation distribution (Ioannidis et al., 2023; Mishra et al., 2022).

Existing research into highly cited scientists has primarily focused on women's representation: what proportion of the elite they form and how it varies across disciplines, countries, and academic-entry cohorts. However, these questions do not address inequalities within the elite. I develop the idea that scientists located above the same high citation threshold (the global citation elite) can actually be in vastly different positions (from the top of the elite to the bottom of the elite). They can also remain in the elite for different periods of time.

In this study, I am interested in what happens to scientists once they cross the citation threshold and reach elite status. I draw a distinction between three important concepts: *inclusion*, *position*, and *persistence*. *Inclusion* means presence in the top 2% annual lists. *Position* shows where scientists are located within the elite. Finally, *persistence* explains whether they appear in the elite once or return to it in subsequent annual elite lists.

These distinctions are especially important in the case of women. Their representation in subsequent annual populations of highly cited scientists may be increasing over time (their participation rate may be rising), yet their numbers may be increasing closer to the top 2% threshold rather than seeing them at the top of the global citation distribution. Women can also appear in the elite but remain in it for a shorter period of time than men. Consequently, more women in the top 2% does not necessarily mean that they have similar positions to men in the elite or similar durability of elite membership.

The two lines of research closest to my study were recently published by Ioannidis et al. (2023) and Mishra et al. (2022). Ioannidis et al. (2023) showed that women were underrepresented in the global top 2% compared with the wider population of Scopus authors. Analyzing subsequent populations of highly cited researchers in 2014–2021, Mishra et al. (2022) showed changes in the gender gap over time and vast differences between disciplines and countries. While these analyses discuss in depth women's representation in the global citation elite (inclusion), my research interest is what happens once male and female scientists are within the global citation elite (position and persistence). I follow the same individuals between subsequent annual lists over time in a longitudinal fashion. I examine where men and women scientists are located after crossing the citation threshold of the top 2% (position) and how long they are able to remain above this threshold (persistence). Some scientists enter the elite and disappear from it, others exit and return after a break, and there are scientists who are permanently present in the elite in every annual list studied.

I used six annual populations of the global top 2% of the most highly cited scientists from 2019–2024. After a data merger, my panel included 465,707 unique scientists and 1,216,089 scientist-year observations (some scientists appeared in more than one annual list). I examined the changes in women's representation, the first observed presence in the data panel, the position in the *c-score* (citation composite index) hierarchy, the total number of annual occurrences in the lists, membership in the durable core, and transitions between presence and absence in subsequent lists. My panel data began in 2019, and from my dataset, it was not known who belonged to the global citation elite before this year. Therefore, the first presence in my panel does not necessarily equate to entering the elite for the first time in one's career.

The contributions of this paper are conceptual and longitudinal. I do not stop at the question of who passes the top 2% threshold—I am also interested in what happens next. Who belongs to the citation elite? Where are scientists located once they pass the threshold? How long do they stay in the elite? These are the three questions of *inclusion*, *position*, and *persistence*. They allow us to test whether women's increasing representation is accompanied by similar changes in their position and in the durability of their elite membership. A higher number (and percentage) of women passing the top 2% citation threshold does not mean that male and female scientists have similar positions above it. It also does not mean that they stay in the elite for an equal number of years.

# 2. Literature review and conceptual frameworks

## 2.1. Inclusion: Women's representation at the top of science

The participation of women in science has increased substantially in the past few decades. However, it varies across countries, disciplines, and career stages (Holman et al., 2018; Huang et al., 2020; Sánchez-Jiménez et al., 2024). More women in science does not necessarily mean their proportionally higher presence in its top echelons (Sugimoto & Larivière, 2023). The more selective the science population examined is, the more important the problem of representation becomes.

In my case, this selective dimension of science is citations. Citations are not a direct measure of research quality, but they are regarded in this research as one of the most important ways to measure scientific visibility and recognition, with implications for academic careers and the availability of research funding. Citation rates are also documented as boosting salaries. Their distribution differs among disciplines, and they depend on collaboration patterns, authorship positions, and field-related citation practices (Aksnes et al., 2019). Male and female scientists are widely reported to differ in citation and self-citation patterns, authorship positions, and locations in collaboration patterns (Sebo & Clair, 2023; Chatterjee & Werner, 2021; Dworkin et al., 2020; Hajibabaei et al., 2023; King et al., 2017; Ross et al., 2022; Wu, 2024).

My analyses are closely aligned to research into highly cited scientists. Using the data of almost 5.8 million authors, Ioannidis et al. (2023) showed that women are considerably less represented in the global top 2% than in the whole population of Scopus authors. They also showed significant differences among disciplines and cohorts. Analyzing highly cited researchers in 2014–2021, Mishra et al. (2022) showed differences between disciplines, regions, and countries. Similar patterns of inequalities among highly cited scientists emerged from research on the Matilda effect in science (Goyanes, 2025).

Previous research has shown that women are underrepresented among highly cited scientists. However, I am interested in the next step: what happens to men and women once they enter the citation elite.

## 2.2. Position: The internal stratification of the citation elite

Membership in the global population of the top 2% is determined by standardized citation indicators developed by Ioannidis and colleagues using the Scopus bibliometric dataset. The foundation of the ranking is the composite *c-score* measure. It includes the total number of citations, the H index, the hm index adjusted to co-authorship, citations to publications connected with specific authorship positions, and others (Ioannidis et al., 2016, 2019, 2020).

The top 2% is a highly selective population, but the differences within this population are still considerable. Citations are generally fairly concentrated. Nielsen and Andersen (2021) have shown that global citation inequality has increased over time, and an ever increasing proportion of citations belongs to a small group of scientists. This concentration does not end with the top 2% threshold. The scientist just above the threshold and the scientist at the very top of the ranking belong to the same citation elite, but they have very different positions within it.

I am interested in whether men and women are similarly located within this selective population. The participation of women in the top 2% may be increasing predominantly near its bottom boundary. At the top of the elite, the changes may be much slower, with a powerful overrepresentation of men. In this case, increasing women's representation in the whole population of the top 2% will tell us little about changes occurring in its top segments.

This is what I term *position*. In this research, I measure position using the relative c-score within annual top 2% populations. I do not treat c-score as a measure of scientific quality; I am interested in something much simpler—do men and women, after crossing the same citation threshold, take similar positions in the internal citation distribution? However, the position in a single year is only a part of the longitudinal picture, as the top 2% ranking is published every year. Consequently, I examine whether the same scientists return in subsequent annual lists or not.

### 2.3. Persistence: The temporal structure of the elite and the accumulation of advantages

Some scientists appear in the top 2% only once. Others appear in the annual lists several times, sometimes after a break. Still others appear in every annual list. In my six-year panel, a difference between one-time presence and repeated presence in the 2% can be observed directly and it is termed *persistence*.

What can be useful in understanding better persistence in membership in elit is the theory of accumulation of advantages. Merton (1968, 1988) showed that earlier recognition makes it easier to obtain future resources, achievements, and recognition. In academic careers, the accumulation of advantages (and, in a parallel manner, disadvantages) includes publications, citations, funding, and visibility. Highly cited scientists in a given year do not start their next year in the list from zero. Their current position is always rooted in earlier careers. At the same time, the accumulation of advantages does influence every scientist in the same way, as shown in recently (Hall & Mairesse, 2026; Mudrak et al., 2026; Siler et al., 2022).

Time is especially important in research into gender inequalities. Women and men may be similarly represented at particular moments of time, but their longer career trajectories may look very different. Existing research has shown gender differences in career development and retention in science (Spoon et al., 2023). Mishra et al. (2022) showed the changing representation of women in subsequent populations of highly cited researchers. In this research, I go a step further, linking the same individuals between subsequent annual lists. I am not interested only in women's representation in a given year; I am also interested in whether the same women and men remain in the top 2% for a similar number of years.

*Persistence* is obviously related to *position*. Scientists located in the c-score distribution further away from the top 2% boundary differ from scientists located just over the threshold. Their current position matters for their presence in the subsequent annual list. However, c-score is not an explanatory variable for gender differences in persistence. The current citation position results from earlier publication and citation history. In addition, earlier presence in the top 2% has its history. I can study how current position and earlier presence in the lists are linked to remaining in the subsequent annual list. However, I cannot refer to causal relationships based on these associations.

I am not able to observe many details of scientific careers with my dataset. I do not have the data about having children, caring responsibilities, or career breaks. I know from previous research that parenthood and caring responsibilities have different consequences for productivity and the careers of men and women (Morgan et al., 2021). They may also have an impact on long-term presence in the citation elite. I am not able to test this with my data, however. Therefore, I return to these processes in the Discussion section.

# 3. Data and methods

## 3.1. Data sources and the construction of the longitudinal panel

I used the publicly available *Updated Science-Wide Author Databases of Standardized Citation Indicators* prepared by Ioannidis and colleagues using the Scopus dataset (Elsevier Repository Data, 2026). The analysis covered six annual single-year datasets from 2019–2024. Each dataset identifies scientists belonging to the global citation top 2% on the basis of disciplinarily standardized citation indicators for a given year.

The challenge in building the data panel was the lack of persistent author identifiers in public files (of the ORCID or Scopus ID type). Therefore, I was not able to directly follow the same individuals between subsequent years. T I constructed a proxy identifier. I combined the lower-case first name and second name of the author (*authfull*) and the year of the first publication indexed in Scopus (*firstyr*):

proxy ID = lower(*authfull*) + *firstyr*

I deduplicated identical identifiers within each year. I obtained 1,216,089 scientist-year observations and 465,707 unique scientists observed at least once in 2019–2024. Depending on the type of analysis, the unit of analysis was scientist, scientist-year, or the transition of the scientist-year from year $t$ to year $t + 1$.

Record linkage on the basis of last names can lead both to false mergers (linking different individuals) and false splits (treating the same person as several individuals). Therefore, I conducted diagnostics and sensitivity analysis of the name-matching procedure. Directly observable collisions of proxy ID, defined as the occurrence of the same proxy ID in more than one record in the same annual source file, were found in 2,223 out of 465,707 identifiers (0.48%) and they were heavily concentrated among last names with Chinese affiliations. I also constructed an alternative identifier using Unicode normalization, diacritic marks, punctuation, and empty spaces in last names.

I subsequently compared the main results about persistence in four specifications: for the proxy ID, for the proxy ID after the exclusion of collision identifiers, for the normalized identifier, and for the normalized identifier and the exclusion of collision cases. The results were practically the same: the unadjusted odds ratio (OR) of female membership in the *durable core* of the elite was in the range OR = 0.841–0.845, and the share of women in the durable core was 18.62–18.72%. The details of the panel, collision analysis, and alternative name-matching procedures are shown in Tables S1-S2 in the Electronic Supplementary Materials (ESM). Sensitivity analyses show that gender differences in persistence are not associated with of the matching procedures applied.

## 3.2. Gender ascription and classification validation

Gender was ascribed probabilistically using genderize.io on the basis of the scientist's first name and the country of affiliation (Wais, 2016). In the baseline analysis, I applied a probability threshold of ≥0.90; cases that did not meet this criterion were classified as *unknown*. With this probability threshold, gender was recognized for 347,193 out of 465,707 scientists (74.55%), including 71,992 women and 275,201 men. Women comprised 20.74% of the population with recognized gender.

I assessed the validity of gender classification in three ways. First, I repeated the main analyses using more restrictive thresholds of ≥0.95 to ≥0.99. The coverage declined by 70.64% (N = 328,992) and 61.92% (N = 288,360), respectively, while the share of women among the classified scientists remained stable (20.58% and 20.94%, respectively, compared to 20.74% at the baseline threshold of ≥0.90). The full sensitivity analyses are presented in Table S3 in the ESM.

Second, I did not treat unknown cases as random missing data. With the probability threshold of ≥0.90, they accounted for 25.45% of the total population, and their frequency differed systematically between countries, fields, cohorts, and persistence levels. The structure of this group is presented in Table S4 in the ESM.

Third, blinded manual validation was completed for 300 scientists from an original stratified sample of 500. The first 300 cases in the randomized order were reviewed. The validator did not know the genderize.io classifications or probability scores. Cases without clear textual evidence from reliable sources were coded as unresolved. I report classification accuracy with a 95% confidence interval, as well as precision, recall, and F1 separately for women and men (see Table S5 in the ESM).

Of the 300 scientists reviewed from the list, 259 were assigned a manual gender label and 41 (13.67%) remained unresolved. Genderize.io agreed with the manual label in 228 of the 259 resolved cases (88.03%; 95% CI: 83.51–91.44). Agreement was 99.07% in the ≥0.99 probability band, 85.42% in the 0.95–0.9899 band, 76.47% in the 0.90–0.9499 band, and 78.85% below 0.90. The errors were asymmetric. Of 130 resolved cases labelled female by genderize.io, 28 were manually labelled male. Of 129 resolved cases labelled male, three were manually labelled female. Precision was 78.46% and recall 97.14% for women; and precision was 97.67% and recall 81.82% for men. At the basic probability threshold of ≥0.90, the reviewed subset contained 242 scientists. Of these, 207 were manually resolved and 187 classifications agreed (90.34%; 95% CI: 85.55–93.66). Female precision and recall were 82.69% and 97.73%; male precision and recall were 98.06% and 84.87%. This subset corresponds to the classification rule used in the main analyses.

## 3.3. Variables and operational definitions

In my analyses, I drew a distinction between three dimensions of membership in the global citation elite: *inclusion*, *position*, and *persistence*. *Inclusion* refers to the presence in an annual top 2% list, *position* shows where a scientist is located within an internal annual citation hierarchy, and *persistence* refers to a single entry or repeated presence in subsequent annual elite lists.

In the case of inclusion, it is important to distinguish between two measures. *First-observed presence* refers to the first scientist's appearance in my 2019–2024 panel. The analysis started in 2020 because in 2019 every scientist, by definition, was observed for the first time. I had no knowledge about who belonged to the top 2% prior to 2019. Therefore, *first-observed* does not necessarily mean entering the citation elite for the first time. *Continuing members* are scientists present in a given year who were observed in at least one previous annual list.

*Annual entry* takes a different perspective. A scientist may be present on the list in the year $t$ but not appear in the list in the year $t - 1$. From 2021 onwards, there may also be individuals who were observed before, then were not present in the next year's list, and then returned to the list. Therefore, annual entry is not the same as *first-observed presence*. There were five such transitions, ending in 2020–2024.

*Position* was measured using the c-score available for each scientist in the source data (Ioannidis et al., 2016, 2019, 2020). Here, I was not interested in the c-score level but in the scientist's position with respect to other elite members of the top 2% in the same year. Therefore, I transformed the c-scores into percentile positions within each annual population. For 2024, I used D1–D10 deciles and permanent percentile positions, scaled in 10-percentile-point units. In the analyses of annual transitions, I used c-score($t$)—that is, the observed position in year $t$ before I knew whether the scientist was present in the list in the year $t + 1$.

*Persistence* is the number of annual lists of the top 2% in which the scientist appeared in 2019–2024. Its value ranges from 1 to 6. The durable core of the citation elite was scientists who were present in five or six lists; the presence in five lists did not have to be continuous. I also tested two more restrictive definitions: at least five subsequent years of presence in the elite, and presence in all six lists. In each case, I discuss only the six observed years. Persistence does not refer to permanent or career-long elite status.

*Annual exit* was observed between two subsequent lists where a scientist was present in the list in year $t$ but was not present in the list in year $t + 1$. The analysis covered five transitions, from 2019 → 2020 to 2023 → 2024. Exit from the list does not imply an exit from science; it means only absence in the subsequent annual list of the top 2%. In the same models, I used *prior elite history*: the number of previous lists in my panel in which the scientist appeared before year $t$.

The remaining variables are more straightforward. *Academic-entry cohort* was determined using the year of the first publication indexed in Scopus (*firstyr*): pre-1990, 1990–1999, 2000–2009, and 2010+. *Broad field* was the most frequently occurring (modal) value of the Science Metrics discipline of a given scientist in the years 2019–2024, while *Country* was the most frequently occurring (modal) observed country of affiliation. In the event of identical numbers of observations, I selected the most recent one (i.e., the most recent country affiliation or the most recent broad field). In analyses focused on countries, I separately show the 15 largest science systems, combining all the other countries into the category of "Other." A full set of definitions is presented in Table S11 in the ESM.

**3.4. Analytical strategy**

I posed four research questions:

**RQ1.** How does women's representation in the global citation elite change in 2019–2024, and are women better represented among the first observed scientists or among continuing members in the panel?

**RQ2.** Do men and women take similar positions in the internal c-score hierarchy of the global citation elite?

**RQ3.** Does persistence of membership in the global citation elite differ between women and men, accounting for academic-entry cohort, country, and broad field?

**RQ4.** Does the association between gender and persistent membership differ between broad fields, academic-entry cohorts, and countries?

I started with six annual populations of the top 2% and changes in their gender composition. For 2020–2024, I examined *first-observed presence*. The logistic regression model included gender and year fixed effects (Mood, 2010); I report ORs with 95% CIs, standardized probabilities, and Average Marginal Effects (AMEs). In 2024, the number of individuals appearing in my panel for the first time was especially high. Therefore, I repeated the model without this year, using the period 2020–2023 (see Table S7 in the ESM). I show *annual entry* separately, and I did not estimate the model. I compared five year-to-year transitions, bearing in mind that from 2021 onwards, among the entrants there may also be scientists returning after a previous absence in the list.

The analysis of positions refers to 2024. I divided my population into 10 deciles according to the internal position of the c-score and examined how the proportion of women changed from decile 1 (D1) to decile ten (D10). Then, I used percentile position as a continuous variable; I provide ORs for transitions by 10 percentile points. Table S9 in the ESM shows the full gender distribution and model results.

In the analysis of persistence of membership, the unit of analysis was the scientist rather than the scientist-year. First, I measured the number of annual lists in which each individual appeared. Then, I separated the durable core of the elite: I defined this as individuals with either five or six appearances in the lists. I estimated the difference between men and women using four logistic regression models. Model M1 includes gender, M2 additionally includes academic-entry cohort, M3 additionally includes country group, and M4 includes also broad field. M4 is my main adjusted model, and I report ORs with 95% CI, standardized probabilities, and AMEs. I do not use the average c-score from the six years as a control variable; it would be measured in the same period as persistence, which means that I would be partially controlling the history of elite membership using a variable emerging from the same history.

I repeated the main analyses using alternative author matching procedures, alternative definitions of persistence, and higher genderize.io probability thresholds. I also examined the scientists classified to the "Unknown" category and how their proportion changed across the population analyzed. The results of these analyses are presented in Tables S1–S11 in the ESM.

# 4. Results

## 4.1. Inclusion: Women's representation and their first-observed presence

The representation of women in the global citation elite increased year-on-year between 2019 and 2024 (Figure 1; Table 1). Among the scientists with a recognized gender, it increased from 18.35% in 2019 to 20.85% in 2024—that is, by 2.50 percentage points. In the same period, the number of women in annual populations of the top 2% increased from 22,698 to 39,041, and the number of men increased from 101,023 to 148,221. The share of *unknown* cases decreased in the same period from 23.22% to 20.27%.

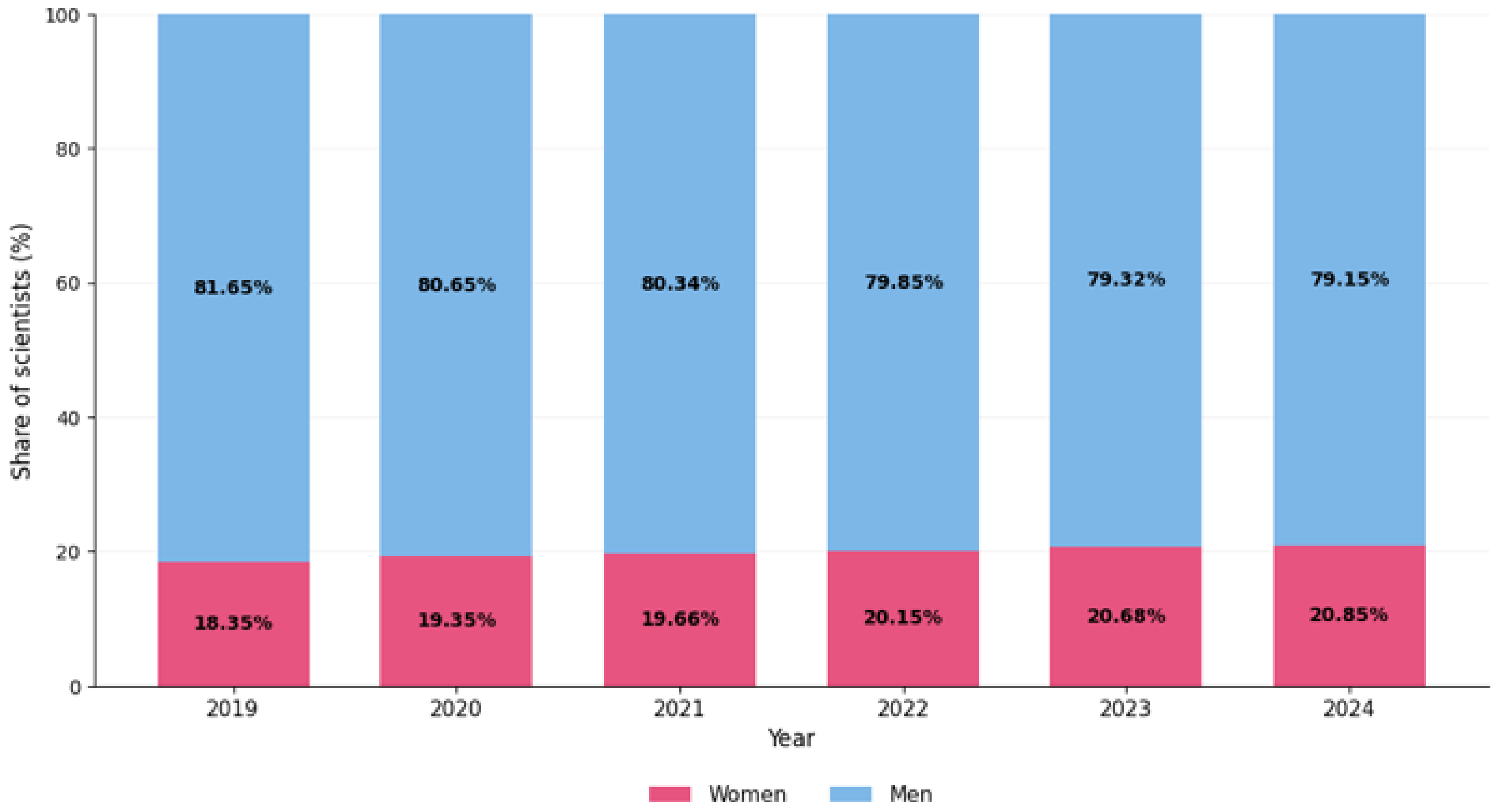

**Figure 1.** Annual gender representation in the global citation elite, 2019–2024. Bars show the proportions of women and men among scientists with known gender in each annual single-year global top 2% list.

**Table 1.** Gender composition of the annual global citation elite, 2019–2024.

| Year | Total, N | Women, N | Men, N | Unknown, N | Known gender, N | Women among known gender (%) | Unknown (%) |
|---|---|---|---|---|---|---|---|
| **2019** | 161,132 | 22,698 | 101,023 | 37,411 | 123,721 | **18.35** | 23.22 |
| **2020** | 189,500 | 28,404 | 118,382 | 42,714 | 146,786 | **19.35** | 22.54 |
| **2021** | 199,440 | 30,453 | 124,444 | 44,543 | 154,897 | **19.66** | 22.33 |
| **2022** | 209,244 | 32,671 | 129,487 | 47,086 | 162,158 | **20.15** | 22.50 |
| **2023** | 221,915 | 35,393 | 135,729 | 50,793 | 171,122 | **20.68** | 22.89 |
| **2024** | 234,858 | 39,041 | 148,221 | 47,596 | 187,262 | **20.85** | 20.27 |
| **Change, 2019–2024** | — | — | — | — | — | **+2.50 pp** | **−2.95 pp** |

*Total N* denotes the number of unique scientists in each annual top 2% list after within-year deduplication using lower(*authfull*) + *firstyr*. Gender was assigned probabilistically using genderize.io based on the first name and country of affiliation. *Known gender* includes scientists classified as female or male with the probability of ≥0.90; all other cases are classified as *unknown*. *Women among known gender* is calculated as Women/(Women + Men), and *Unknown (%)* is relative to *Total N*. pp = percentage points.

The representation of women increased with each subsequent year, but in 2024, women still only comprised 20.85% of scientists with a recognized gender. The unknown group remained large, although its proportion decreased from 23.22% in 2019 to 20.27% in 2024. Proportions of men and women were calculated for the population with recognized gender. I left unknown as a separate category and show its magnitude in each year.

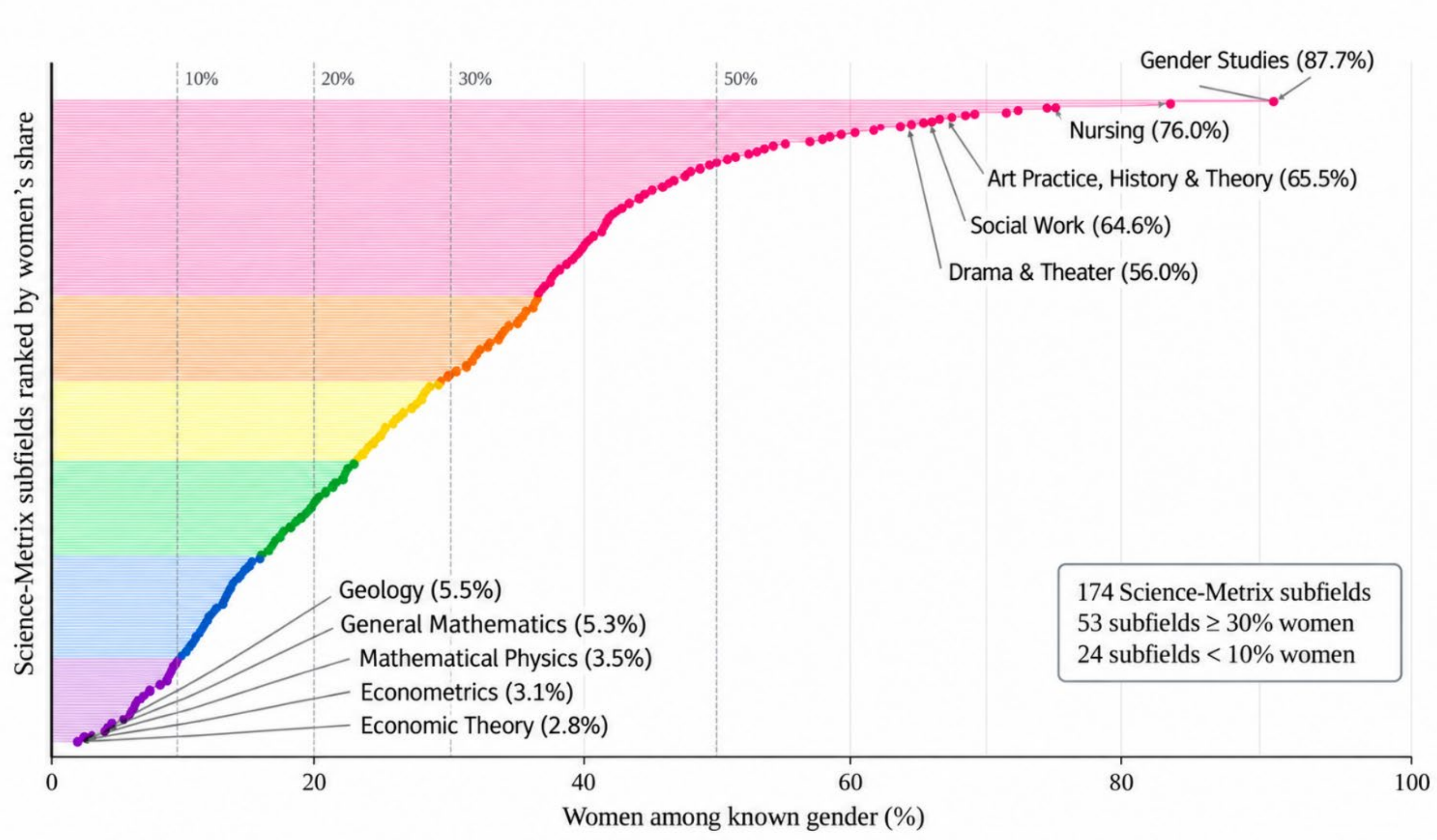

**Figure 2.** Women's representation across 174 Science-Metrix subfields in the global citation elite, 2024. Subfields are ranked from the lowest to the highest proportion of women among scientists with known gender in the 2024 global top 2%. Dashed lines mark 10%, 20%, 30%, and 50%. The five subfields at each end of the distribution are labeled.

The share of women differed substantially among the 174 Science-Metrix subfields (Science-Metrix, 2018). (Figure 2). The range was significant: from 2.8% in Economic Theory to 87.7% in Gender Studies. In Econometrics, 3.1% of represented scientists were women, in Mathematical Physics 3.5%, in General Mathematics 5.3%, and in Geology 5.5%. In contrast, in Nursing, 76.0% of the represented scientists were women; in Art Practice, History & Theory 65.5%; in Social Work 64.6%; and in Drama & Theater 56.0%. In the majority of subfields, however, the share of women was much lower. In only 53 out of 174 (30.5%) subfields was their share higher than 30%, and in 24 (13.8%) it did not reach 10%.

Women appeared for the first time more often than men in my panel (Table 2). In 2020, this was the case for 41.48% of women and for 36.43% of men, a difference of 5.06 pp. In subsequent years, the difference was smaller, but it always showed the same direction: 3.09 pp in 2021, 2.67 pp in 2022, 2.84 pp in 2023, and 2.66 pp in 2024; the OR for women was higher than 1.0 in each year (1.120–1.237).

For the five years pooled, the result was straightforward. In the model with year fixed effects, the OR for women was 1.178 (95% CI: 1.164–1.193), equivalent to a standardized probability of 29.73% for women and 26.54% for men, with a difference of 3.19 pp.

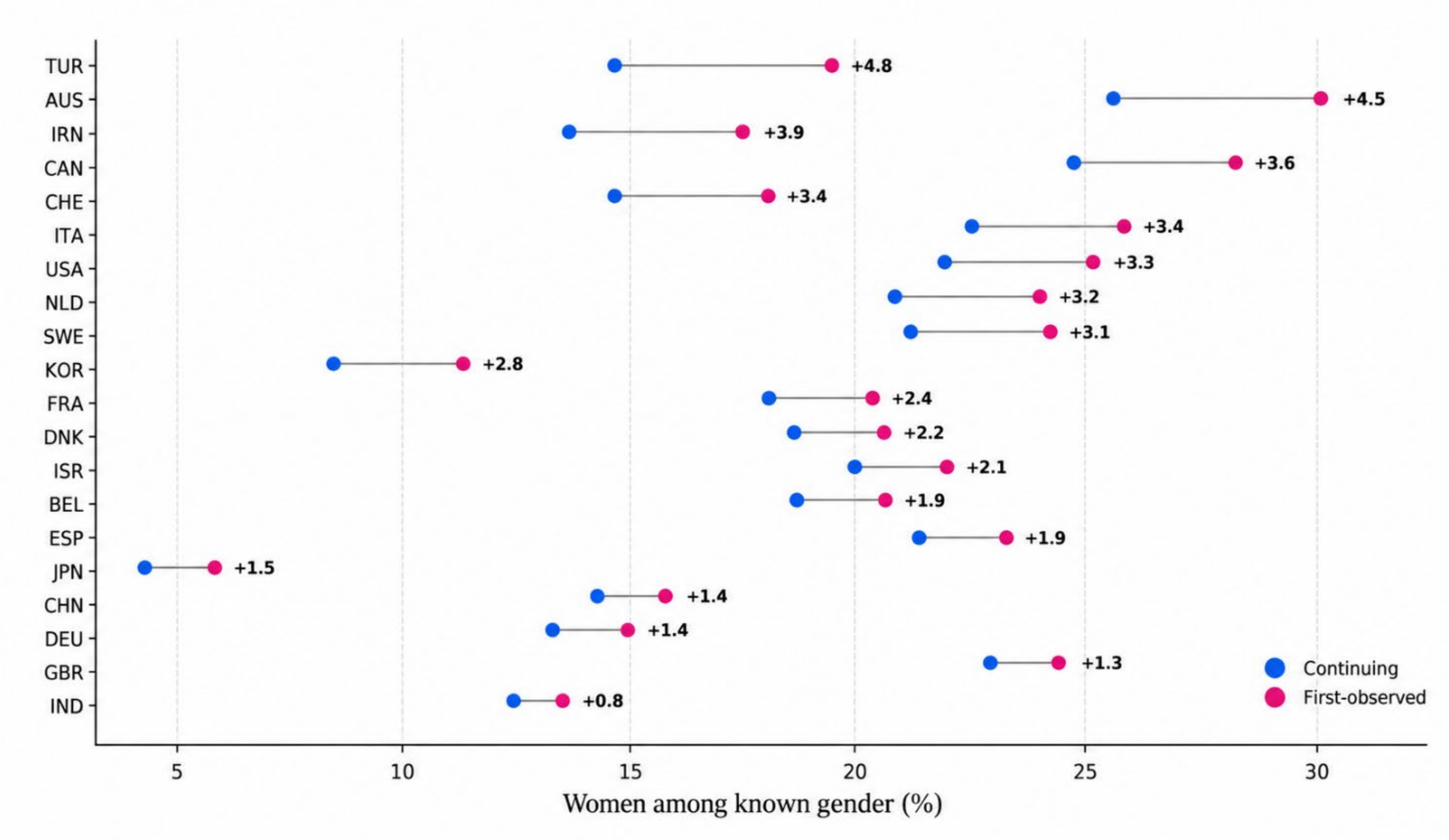

**Figure 3.** Women among first-observed scientists (in red) and continuing members (in blue) in the 20 largest national citation-elite populations, 2020–2024. Shares refer to scientists with known gender only and are pooled across 2020–2024. First-observed scientists had not appeared in an earlier list in my 2019–2024 panel. Labels show the difference between continuing members and first-observed members in percentage points.

In all 20 countries, the proportion of women was higher among first-observed scientists than among continuing members (Figure 3). However, the differences were not large, ranging from +0.8 pp in India to +4.8 pp in Turkey. Australia showed a similar difference (+4.5 pp), and this was followed by Iran (+3.9 pp), Canada (+3.6 pp), Switzerland and Italy (+3.4 pp), and the United States (+3.3 pp).

**Table 2.** First-observed presence in the annual global citation elite by gender, 2020–2024. Gender is classified using genderize.io with a probability threshold of ≥0.90. Men are the reference category (OR = odds ratio; CI = confidence interval; pp = percentage points).

| **Year** | **Known gender, N** | **First-observed, N** | **Men first-observed (%)** | **Women first-observed (%)** | **Difference (pp)** | **OR for women** | **95% CI** |
|---|---|---|---|---|---|---|---|
| **2020** | 146,786 | 54,907 | 36.43 | **41.48** | **+5.06** | **1.237** | 1.205–1.270 |
| **2021** | 154,897 | 34,706 | 21.80 | **24.88** | **+3.09** | **1.188** | 1.154–1.224 |
| **2022** | 162,158 | 33,896 | 20.37 | **23.03** | **+2.67** | **1.170** | 1.137–1.205 |
| **2023** | 171,122 | 30,648 | 17.32 | **20.16** | **+2.84** | **1.205** | 1.170–1.241 |
| **2024** | 187,262 | 69,315 | 36.46 | **39.12** | **+2.66** | **1.120** | 1.094–1.146 |
| Pooled, 2020–2024 | 822,225 | 223,472 | 26.54 | 29.73 | +3.19 | 1.178 | 1.164–1.193 |

In Table 2, first-observed presence refers to the scientist's first appearance in the 2019–2024 panel, and it does not necessarily indicate their first entry into the global citation elite in their career. The year 2019 was excluded from my analysis because all scientists present in the first year of the panel were automatically observed for the first time. Annual percentages show the share of first-observed scientists among women and men present in a given annual list. Annual ORs are for women compared with men in each year. The pooled estimation is based on a logistic regression for the years 2020–2024 with year fixed effects (Mood, 2010). Pooled N is the number of scientist-year observations rather than the number of unique scientists. Pooled percentages are standardized expected probabilities, and the difference between them is shown as an average marginal effect (AME).

In each year between 2020 and 2024, women showed a higher probability of their first appearance in the panel than men. In the pooled model with year fixed effects, the OR was 1.178 (95% CI: 1.164–1.193), and the standardized probability was 29.73% for women and 26.54% for men, equivalent to an AME of +3.19 pp.

Because the population of the first-observed was atypically large in 2024 (69,315 scientists with recognized gender, compared with 30,648 in 2023), the model was also estimated for the years 2020–2023. The exclusion of the year 2024 did not weaken the results: OR = 1.202 (95% CI: 1.185–1.219), with standardized probabilities of 26.97% for women and 23.61% for men and an AME of +3.36 pp (see Table S7 in the ESM).

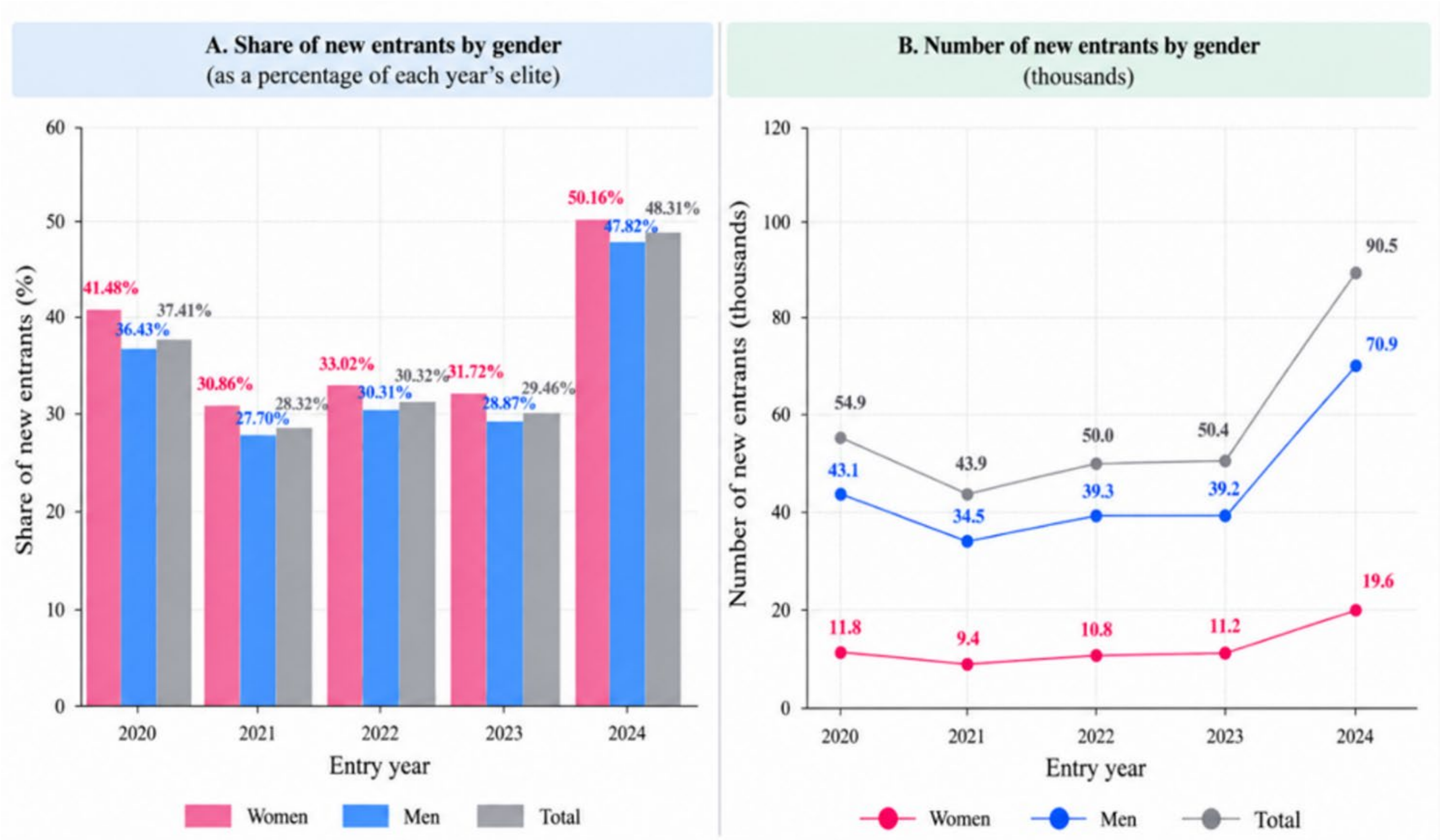

**Figure 4.** Entry into the global citation elite, 2020–2024. Panel A shows annual entry rates into the global citation elite (top 2%) by gender and for the whole population. An entry in year *t* denotes a scientist present in the annual top 2% list in year *t* who is not present in the list directly preceding, in the year *t−1*. Percentage values were calculated with reference to the relevant elite population in the year *t*. Panel B shows the relevant numbers of individuals entering the elite (in thousands).

The analysis began in 2020 because it was not possible to observe elite membership prior to 2019. Annual entry represents the measure of transition from year to year, and it differs from first-observed presence. From 2021, among the entrants, there could also be scientists observed earlier in the panel who, after a period of absence, returned to the list. Gender was classified as male or female using genderize.io, with a probability threshold of ≥0.90.

Annual entry rates were higher for women than for men in all the years analyzed (Figure 4). For women, they ranged from 30.86% to 50.16%, and for men, they ranged from 27.70% to 47.82%. In 2024, the total entry rate rose substantially to 48.31%, with 90,464 scientists entering the elite.

My results indicate a gradual increase in the representation of women in the annual citation elite and their relatively stronger representation among individuals appearing for the first time in the panel. Because of the left boundary of observations in 2019, first-observed presence does not identify the first entry into the global top 2% in a scientist's career. The result refers to inclusion within the observed panel rather than the first-ever entry into the citation elite.

## 4.2. Position: The position of women within the internal c-score hierarchy

In 2024, women were not equally represented internally within the top 2% (Figure 5). In the lowest *c-score* decile (D1), women comprised 26.88% of the scientists with a recognized gender. In the highest decile (D10), women accounted for 15.23%. The difference between the two extremes of this distribution was 11.65 pp.

The decline was not even between subsequent deciles. In several deciles, the percentage of women slightly increased. However, the direction of the whole distribution was clear. In the continuous model, a 10-percentile-point increase in c-score was equivalent to an OR of 0.951 (95% CI: 0.947–0.954) for women. The full results for D1–D10 and the results of the model are presented in Table S9 in the ESM.

What I measure here is the position within the top 2% *c-score*, which in this analysis is not a measure of scientific quality. I also do not describe the relationship between gender and *c-score* in a causal manner. I am interested in a much simpler question: where are men and women located after crossing the same top 2% threshold?

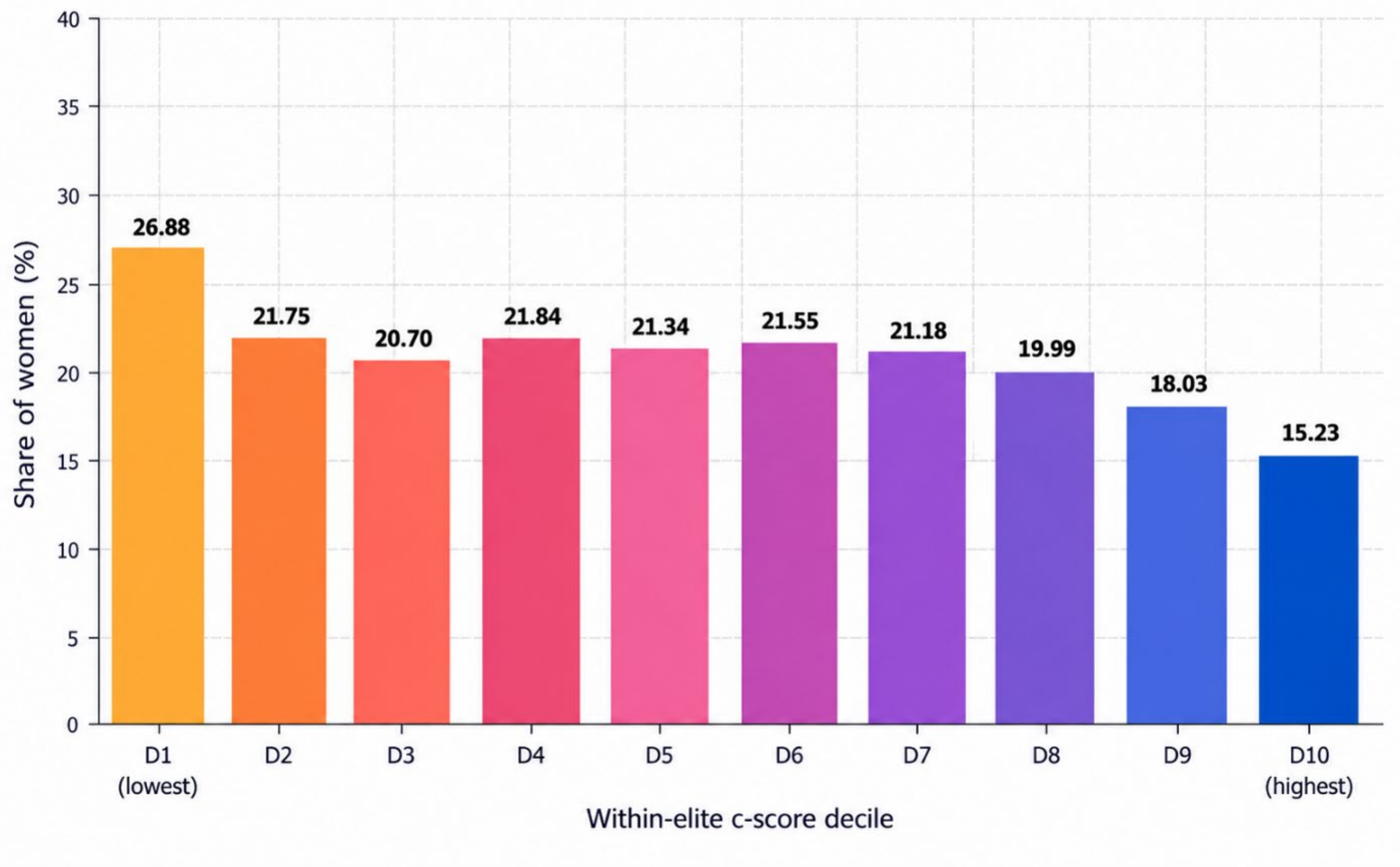

**Figure 5.** The representation of women by their *c-score* position within the global citation elite, 2024.

Figure 5 shows the representation of women among scientists with recognized gender in particular deciles of the distribution of the *c-score* of the top 2% in 2024. D1 denotes the lowest decile, and D10 denotes the highest decile. Gender was classified using genderize.io, with a probability threshold of ≥0.90.

The proportion of women clearly changed with the *c-score* position (Figure 5). In D1, women made up 26.88%, and in D10 they accounted for only 15.23%, a difference of 11.65 pp. The decline was not even between subsequent deciles, but its direction was straightforward. The same result was obtained for the continuous position of *c-score*. A 10-percentile-point increase in *c-score* position was equivalent to an OR of 0.951 (95% CI: 0.947–0.954) for women.

## 4.3. Persistence and annual exit from the citation elite

Differences were also observed when time was introduced into the analysis (Figure 6A). Women accounted for 21.58% of scientists present only in one annual list and 17.69% of scientists who were present in all six lists. For the durable core, the difference was similar; 21.20% of women (15,261/71,992) and 24.24% of men (66,698/275,201) belonged to this category. The difference was 3.04 pp (Table 3, Panels A–B).

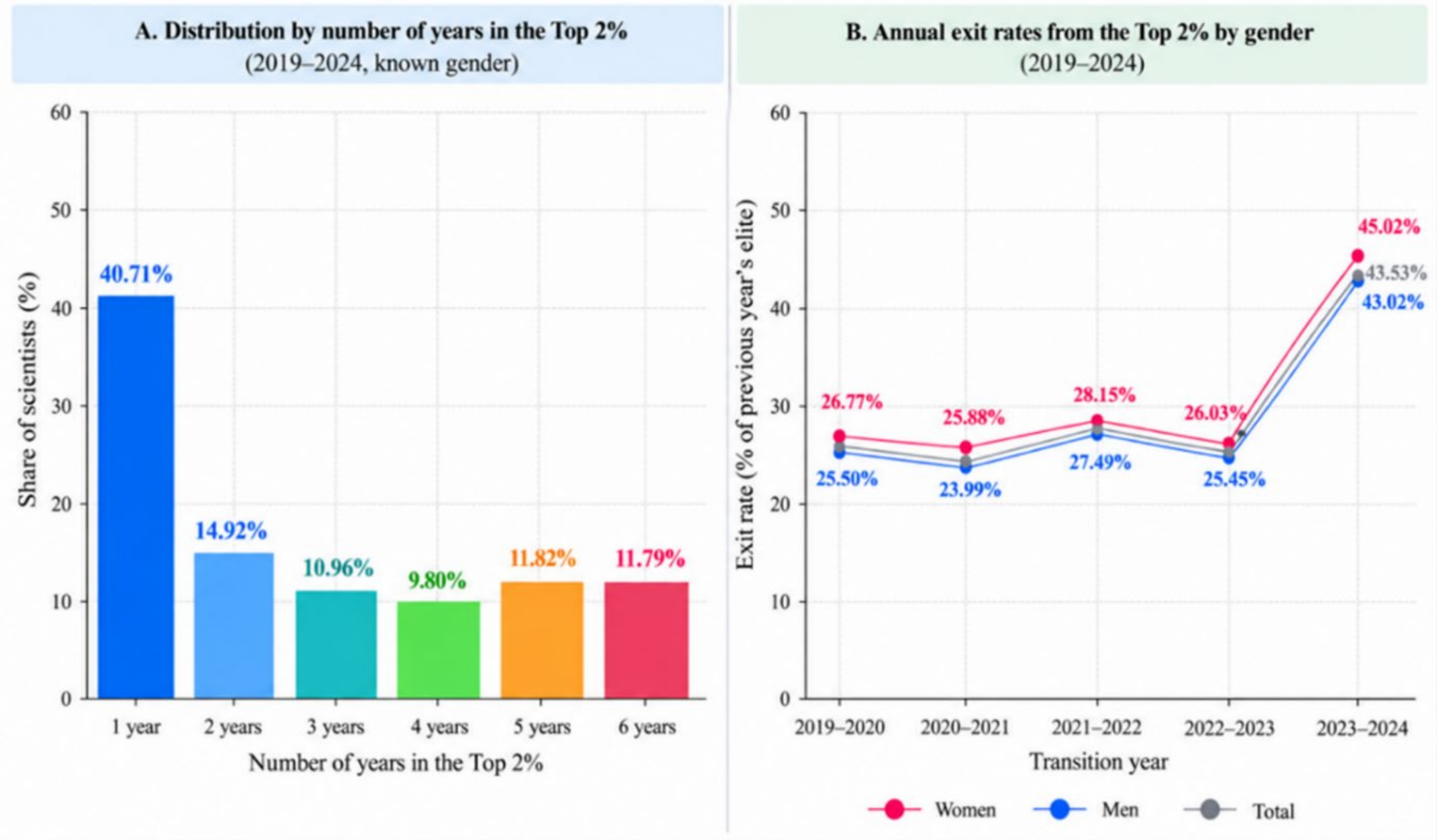

**Figure 6.** *Persistence in elite membership* and annual exit from the global citation elite, 2019–2024. Panel A shows the number of annual top 2% lists in which scientists with recognized gender appeared in 2019–2024 ($N = 347,193$). Panel B shows unadjusted annual exit rates for women and men in five transitions from 2019 → 2020 to 2023 → 2024. *Annual exit* is presence in year $t$ followed by absence in year $t + 1$. Gender was classified using genderize.io, with a probability threshold of ≥0.90.

The majority of scientists were not present in the top 2% lists in all six years (Figure 6A), with 40.71% of scientists appearing only once and 11.79% appearing in all six years. Annual exit (Figure 6B) showed a different pattern. In the first four transitions, exit rates for women and men were at the level of 24–28%. In the 2023→2024 transition, they suddenly increased to 45.02% for women and to 43.02% for men. In all five transitions, raw rates were slightly higher for women than for men.

In the models, this difference was clearly smaller (see Table S8 in the ESM). It substantially declined after the current c-score($t$) was included, and after prior elite history was added, it became invisible. In the fully adjusted model, the difference was small and reversed.

**Table 3.** Gender differences in persistence and durable-core membership in the global citation elite, 2019–2024.

**Panel A.** Distribution by number of annual top 2% appearances.

| Annual appearances | Women, N | Men, N | Known gender, N | Women (%) | Difference from 1-year membership (pp) |
|---|---|---|---|---|---|
| 1 year | 30,500 | 110,844 | 141,344 | 21.58 | Reference |
| 2 years | 11,177 | 40,639 | 51,816 | 21.57 | −0.01 |
| 3 years | 7,957 | 30,092 | 38,049 | 20.91 | −0.67 |
| 4 years | 7,097 | 26,928 | 34,025 | 20.86 | −0.72 |
| 5 years | 8,019 | 33,012 | 41,031 | 19.54 | −2.03 |
| 6 years | 7,242 | 33,686 | 40,928 | 17.69 | −3.88 |
| Total | 71,992 | 275,201 | 347,193 | 20.74 | — |

**Panel B.** Durable-core membership by gender.

| Gender | Total, N | Durable core, N | Durable-core membership (%) |
|---|---|---|---|
| Women | 71,992 | 15,261 | 21.20 |
| Men | 275,201 | 66,698 | 24.24 |
| Total | 347,193 | 81,959 | 23.61 |
| Women − men difference | — | — | −3.04 pp |

**Panel C.** Logistic regression models predicting durable-core membership.

| Model | Specification | Analysis pop., N | OR for women | 95% CI | Adjusted P: Women (%) | Adjusted P: Men (%) | AME (pp) |
|---|---|---|---|---|---|---|---|
| **M1** | Gender | 347,193 | 0.841 | 0.824–0.858 | 21.20 | 24.24 | −3.04 |
| **M2** | Gender + academic-entry cohort | 347,193 | 0.864 | 0.847–0.882 | 21.59 | 24.13 | −2.54 |
| **M3** | M2 + country group | 347,193 | 0.871 | 0.853–0.888 | 21.70 | 24.10 | −2.39 |
| **M4** | M3 + broad field | 347,168 | 0.886 | 0.868–0.905 | 21.95 | 24.03 | −2.08 |

Persistence represents the number of annual top 2% lists in which the scientist appeared in the years 2019–2024. The durable core includes scientists present in five or six annual lists; for five lists, the presence did not have to be continuous. In Panel A, the differences were calculated relative to scientists present in only one annual list. Panel C presents the logistic models for durable-core membership. The model M1 includes gender; M2 additionally includes academic-entry cohort, M3 adds country group, and M4 adds broad field to the model. The reference category in the models is men. The sample in M4 was smaller by 25 individual cases because of the missing information about their broad field (OR = odds ratio; AME = average marginal effect; CI = confidence interval; pp = percentage points).

Women's representation declined with the number of annual appearances (Table 3). Of the scientists present in the list only once, women comprised 21.58%; among scientists present in all six years, women accounted for 17.69%; 21.20% of women and 24.24% of men belonged to the durable core, a difference of 3.04 pp.

Part of this difference remained after adjustment of the composition of the population. The OR for women was 0.841 in model M1, 0.864 after adding academic-entry cohort (M2) and 0.871 after adding country group (M3). In M4, which also included broad field, OR = 0.886 (95% CI: 0.868–0.905). The predicted probabilities were 21.95% for women and 24.03% for men. The difference was −2.08 pp (95% CI: −2.43 to −1.72).

Transitions between two annual lists differed substantially (Table S8 in the ESM). In the first model, women had slightly higher chances of annual exit (OR = 1.064; AME = +1.28 pp). After the current c-score(t) was added, the OR decreased to 1.022 and the AME decreased to +0.44 pp.

When elite membership history was introduced to the model, the estimate changed, with OR for women being 0.990 (95% CI: 0.977–1.004). Each additional year of prior presence in the top 2% was associated with smaller chances of later exit from the elite (OR = 0.617). In the fully adjusted model, the OR for women was estimated to be 0.969 (0.956–0.983), and the AME was −0.58 pp (−0.84 to −0.32). The women × c-score(t) interaction was very small: OR = 1.016 (1.001–1.032; p = 0.038).

The two results should be kept separate. Women have smaller chances to belong to the durable core of the elite. But when controlled for the current position and earlier elite membership history, they also did not have higher chances of dropping out from the next annual list. The gender persistence gap resulted from membership history (and not from gender difference in annual exits).

### 4.4. Heterogeneity across fields, cohorts, and countries

The gender gap in durable-core membership differed across the citation elite. Adding interactions, M4 showed the differences by broad field, academic-entry cohort, and country group (LR $\chi^2 = 54.73$, df = 19, $p < 0.001$; LR $\chi^2 = 72.06$, df = 3, $p < 0.001$; and LR $\chi^2 = 28.31$, df = 15, $p = 0.020$, respectively). Differences were stronger for entry cohorts and broad fields and weaker for countries. Figure 7 shows the gap by broad field.

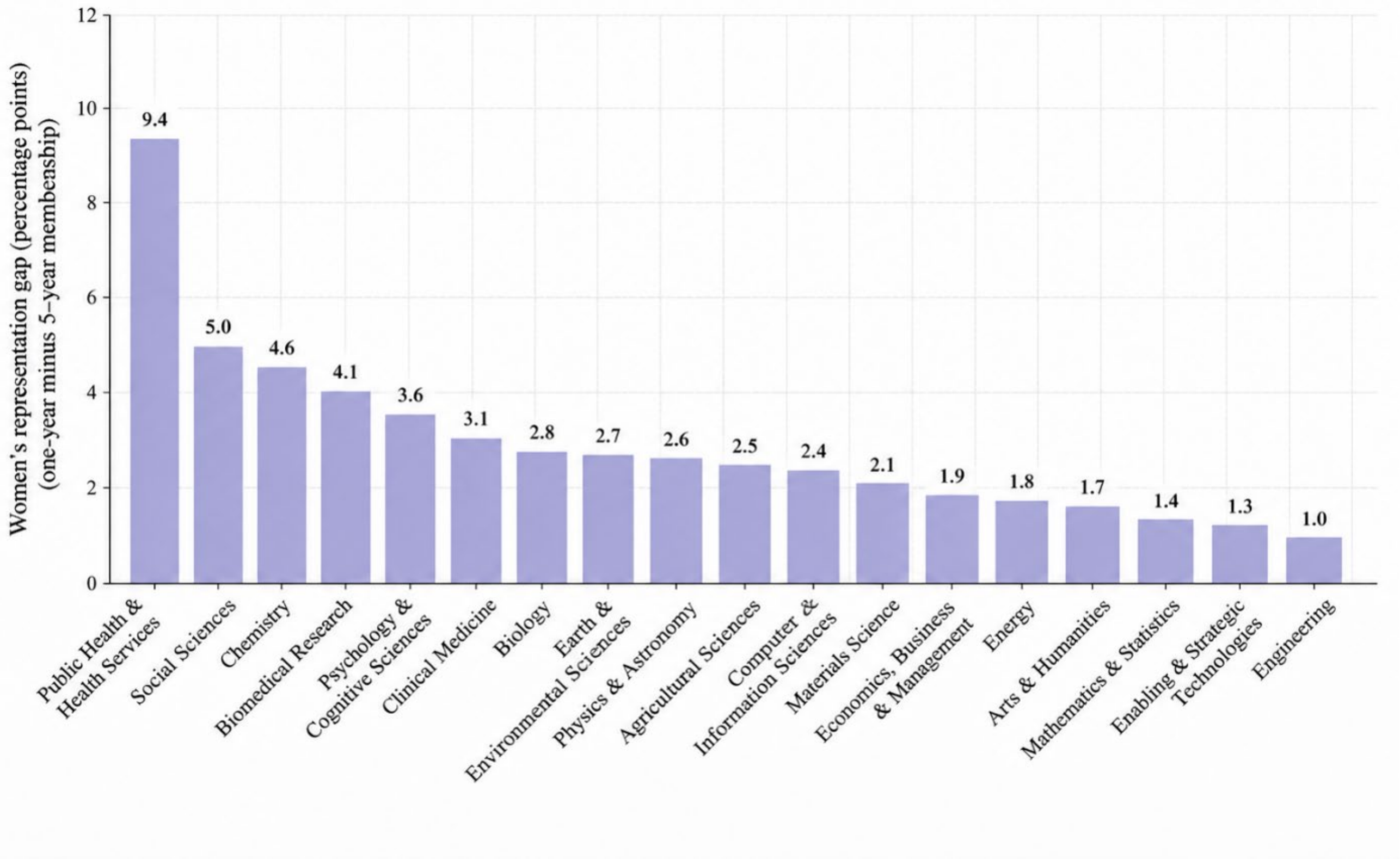

**Figure 7.** Gender gap in durable-core membership by broad field, 2019–2024. The gap is the difference in representation in scientists present in only one annual list and scientists present in five or six annual top 2% lists. Positive values mean higher women's representation among one-year members than in the durable core. Broad fields are ordered by the size of the gender gap. Values are unadjusted and refer to scientists with known gender.

The difference between one-year members and the durable core varied considerably between broad fields (Figure 7). It was largest in Public Health & Health Services (+9.4 pp). In Social Sciences, it was +5.0 pp, in Chemistry +4.6 pp, in Biomedical Research +4.1 pp, and in Psychology & Cognitive Sciences +3.6 pp. At the other extreme was Engineering (+1.0 pp), Enabling & Strategic Technologies (+1.3 pp), and Mathematics & Statistics (+1.4 pp). All differences are unadjusted.

The field differences remained after the adjustment. In Public Health & Health Services, the AME was −4.90 pp (95% CI: −6.71 to −3.09), in Chemistry −4.76 pp, and in Biomedical

Research −4.33 pp. In Clinical Medicine, despite the size of the field, the difference was much smaller: −0.86 pp (−1.38 to −0.33).

The academic-entry cohort results also differed. In the pre-1990 cohort, the probabilities of durable-core membership were almost the same for men and women: 26.49% and 26.79% (AME = −0.29 pp; OR = 0.985). In the 1990–1999 cohort, the difference was −1.85 pp. The difference was largest for the 2000–2009 cohort: 20.04% for women and 23.46% for men, or −3.43 pp (OR = 0.815). Finally, for the 2010+ cohort, the difference was −2.77 pp.

I do not view this as a simple trend between subsequent generations of scientists. The cohort was associated with career length and with the time the scientist had prior to my six-year observation window.

Country differences were smaller. Among the largest science systems, AME ranged from −4.36 pp in South Korea (OR = 0.714) to +0.03 pp in the United Kingdom (OR = 1.002). For some countries, confidence intervals included zero.

Academic-entry cohort does not show directly at which career stage scientists were observed in the years 2019–2024. Therefore, I show the elite composition by academic age separately (Figure 8).

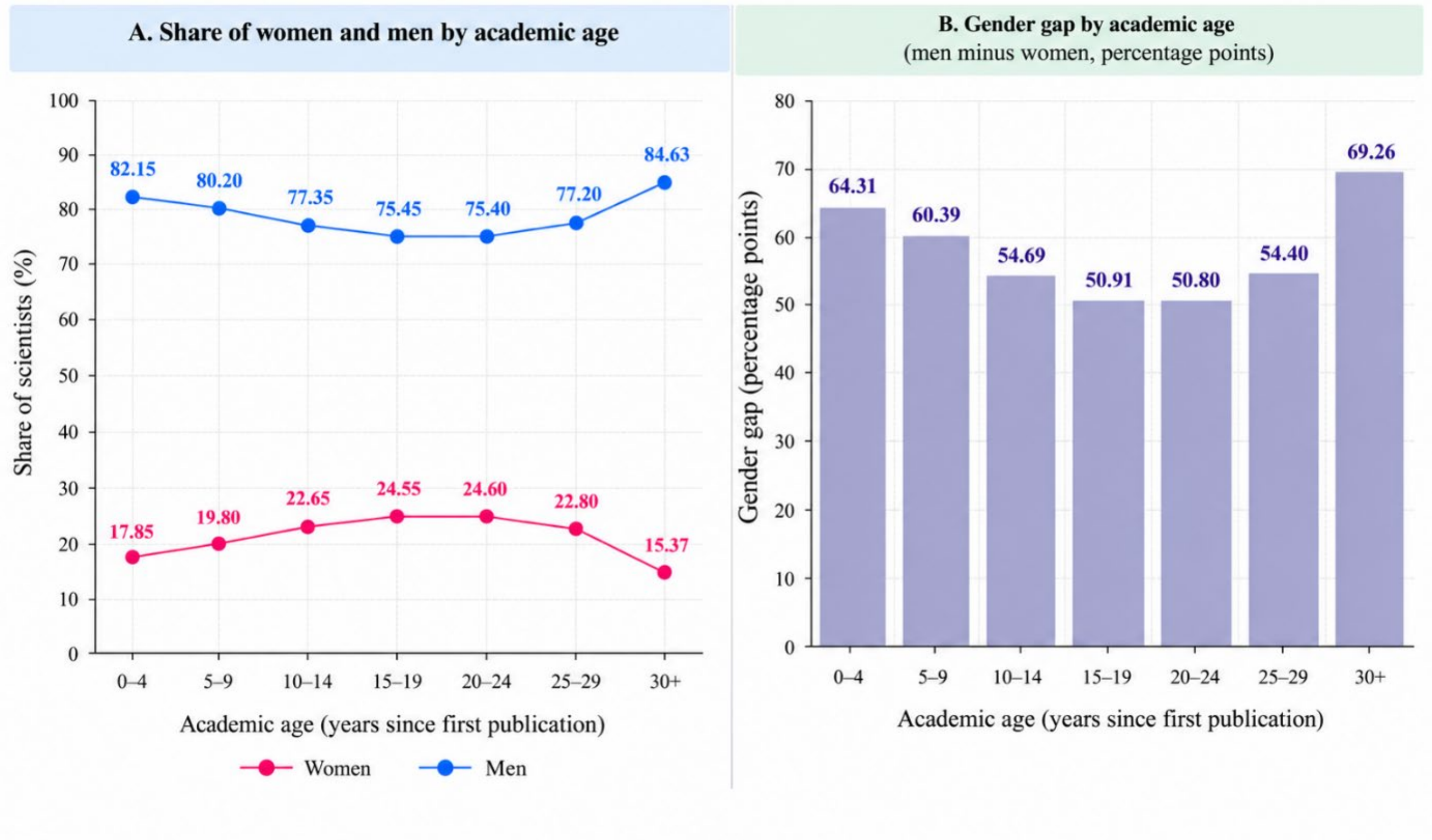

**Figure 8.** The gender structure of the global citation elite by academic age group, 2019–2024.

Panel A presents the percentages of men and women among scientists with classified gender by academic age group. Panel B presents the corresponding difference between men and women. Academic age is the observation year minus *firstyr*. Values pool scientist-year observations from 2019–2024. Women's representation did not change linearly with academic age (Figure 8). It was 17.85% at 0–4 years, rose to 24.60% at 20–24 years, and then fell to 15.37% at 30+ years. The gender gap narrowed correspondingly from 64.31 pp to 50.80 pp and then widened to 69.26 pp. Thus, women were most strongly represented in the middle of the academic age distribution.

### 4.5. Robustness

The main results of this research remained stable under alternative measurements and sample specifications. First, the gender gap in durable-core membership was basically unchanged under four different author-matching procedures. Unadjusted ORs for women were in a very narrow range of 0.841–0.845, and the share of women in the durable core was between 18.62% and 18.72% (see Tables S1–S2 in the ESM). The results show that persistence did not depend on the basic linkage procedure of lower(*authfull*) + *firstyr*.

Second, the fully adjusted persistence gap remained visible when alternative definitions of durable membership were used. The fundamental definition of five or six appearances in the elite resulted in an OR of 0.886 (95% CI: 0.868–0.905; AME = −2.08 pp). The requirement of at least five subsequent annual appearances resulted in an OR of 0.892 (0.872–0.912; AME = −1.78 pp), and the requirement of presence in all six lists resulted in an OR of 0.851 (0.828–0.876; AME = −1.57 pp) (Table S6 in the ESM).

Third, increasing the genderize.io gender classification threshold from ≥0.90 to ≥0.95 and ≥0.99 decreased the coverage of gender classification from 74.55% to 70.64% and 61.92%, respectively, with a basically stable share of women among scientists with a classified gender (20.74%, 20.58%, and 20.94%; Table S3 in the ESM). Unknown cases were not treated as random missing data (Table S4 in the ESM).

Finally, first-observed presence remained stable after the removal of an atypically large population of first-observed scientists in 2024 (OR = 1.202, 95% CI: 1.185–1.219; AME = +3.36 pp; Table S7 in the ESM). The tests showed that inclusion and persistence patterns remained stable when alternative operationalizations and specifications of the sample were used.

## 5. Discussion

### 5.1. From representation to stratification: Inclusion, position, and persistence

Previous research shows that women are underrepresented among highly cited scientists and that the size of this gap varies considerably between fields, countries, and cohorts (Ioannidis et al., 2023; Mishra et al., 2022). My data show that the representation of women in the top 2% remains low but is steadily increasing. Women more often than men appeared for the first time in my panel.

Once the threshold for the global citation elite is crossed, a new pattern emerges. Women comprise 26.88% of the lowest c-score decile (D1) and only 15.23% of the highest decile (D10). The difference is substantial. Entering the elite thus does not tell us much about the position taken within this population. This matters for the interpretation of the increase in women's representation. In 2019–2024, the number and percentage of women in the top 2% increased, but women were not similarly represented in the whole c-score distribution, being much more often closer to the bottom boundary of the top 2% than to its top. If I focus only on the general representation of women in the elite, I cannot see this difference.

Another problem emerges when time is factored in. Women constitute 21.58% of scientists present only in one annual list and 17.69% of scientists present in all six lists. Under the basic definition of the durable core, the gap remains after adjustment for academic-entry cohort, country, and discipline (OR = 0.886; AME = −2.08 pp). The difference does not disappear under more restrictive definitions of persistence. It also does not essentially depend on alternative name-matching procedures.

It could be expected that the lower share of women in the durable core results simply from their more often dropping out of subsequent annual lists. The results are not so clear cut, however. Raw annual exit rates are slightly higher for women than for men in all five transitions. After adjustment for the current c-score, the difference clearly decreases, and after adjustment for prior elite membership history, it practically disappears. In the fully adjusted model, the difference is small and reversed (OR = 0.969; AME = −0.58 pp). The women × c-score($t$) interaction is also very small (OR = 1.016).

Persistence and annual exit rates therefore tell us different things. The annual exit rate refers to a single transition between two subsequent annual lists. Persistence refers to five or six years of presence in the lists. Scientists enter subsequent transitions with different prior elite membership histories and from different positions in the c-score distribution. When these differences are taken into account, I do not observe a bigger annual exit rate for women. This does not mean that c-score and prior elite membership history causally explain the gender gap in persistence. Both variables are already parts of a scientist's previous career history.

There are three simultaneously occurring processes: women's representation in the top 2% is increasing; women are less strongly represented in the upper parts of the internal c-score distribution; and women are less strongly represented among the most persistent members of the elite. An increase in representation does not explain the other two ongoing processes. Therefore in my analyses, I clearly distinguish between inclusion, position, and persistence.

## 5.2. Heterogeneity and possible mechanisms

The gender gap in persistence in elite membership differs across the population studied. The biggest differences are observed between broad fields and academic-entry cohorts; country differences are much smaller. A pooled estimate is useful in describing the whole population, but it does not show this variation.

A STEM versus non-STEM divide is not observed. There are large gaps in fields that have little in common, and in other fields, the gaps are small. There are substantial differences between Public Health & Health Services or Biomedical Research and Clinical Medicine or Engineering. The reduction of these patterns to a single field characteristic is not useful.

Doing science looks different in different fields. Fields vary considerably in publication intensity, team size, collaboration patterns, authorship conventions, and the pace of citation accumulation. Gender differences in some fields are well documented (Dworkin et al., 2020; Hajibabaei et al., 2023; Ross et al., 2022; Wu, 2024). The gender gap differs across fields, but it cannot be seen what produces these differences.

Cohort results are interesting to analyze. In the case of scientists who began publishing before 1990, the adjusted difference between men and women is very small, and it is bigger for later cohorts. This is not evidence that gender inequality increased in successive generations of scientists. A scientist representing the pre-1990 cohort and a scientist representing the 2010+ cohort are in entirely different points of their careers, even though I observe both in the same years 2019–2024.

Selection also matters. Women who started publishing prior to 1990 and three decades later belong to the global top 2% are a very specific group. Most women starting their careers in science at that time never reached that stage or did not remain scientifically active for so long. The same is true of men, but their selection would have taken a different course. Survivors in science can be seen rather than the initial population of these women. With merely six years of observation, distinguishing between cohort, career stage, period, and selective survival is impossible.

Family and career development issues are also sensitive areas of inquiry. Previous research shows that parenthood, caring obligations, and breaks in careers have different influence on the careers and productivity of women and men (Morgan et al., 2021). A break in career does not have to stop entering the top 2% at some point, but it can influence one's future continuous presence in it. However, the data about children, parental leaves, and caring obligations are not available in our dataset. Surveys, interviews, and institutional or national administrative data would be especially useful in this context.

Country differences are smaller, and more difficult to interpret. Science systems have different employment and family policies. My country variable does not measure any of these; it is the modal (dominating) country of affiliation in the period 2019–2024. It does not show where a scientist was taking their education, where they were working before, or whether they were changing country in the course of their career. In my data, I also have an unequal coverage of genderize.io, especially in East Asia.

Cumulative advantage theory provides a useful context to my results. Scientists do not appear in each subsequent year with a clean slate. They have prior publications, citations, collaborations, and resources in their individual career histories. They enter the elite and annual exit models show the meaning of these histories. The initial gender difference diminishes when the current c-score and the prior elite history are considered, and it is small and reversed after full adjustment.

However, this is also where the data used reach their limit. C-score(*t*) and prior elite history do not appear from nowhere - they result from previous careers. Applying them in the model, I compare women and men with similar current positions and comparable history of presence in the top 2%. But I do not learn why these histories became different in the first place.

Therefore, I interpret the interactions found narrowly. They show that the gender gap differs in size between fields, cohorts, and countries. However, they do not show the causal relationships. The data show where the gaps appear and how they change with differences in prior elite history, but they do not show how that history was produced in the first place.

## 5.3. Implications and limitations

My results refer to women who belong to the global citation elite, but the problem I discuss is much wider. Scientific elites are produced using certain thresholds: citations, publication productivity, publishing in prestigious journals, financing, or awards. The threshold shows who is inside the selected group without showing the differences among individuals inside the group.

My analyses clearly show that women's representation in the top 2% is rising. If I stopped my analysis at this level, the conclusion would be simple: the gender gap in the global citation elite is gradually decreasing. However, above this threshold, men and women take different positions in the c-score distribution and they differ in persistence in staying in the elite. At the same time, however, lower persistence of women does not translate easily into a higher annual exit rate. These differences canot be captured by a single measure of representation.

The same problem may arise with other definitions of scientific elites. I can know what proportion of women belongs to the class of the most productive scientists, obtains the largest research grants, and publishes in prestigious journals, but there still remains a question of where women are located within these selective groups and whether they stay in these groups as long as men do. I do not know whether the results of such studies would be similar to ours. However, the application of such notions as *inclusion*, *position*, and *persistence* provides a useful way to pose such questions.

My research is not without limitations. The most important limitation is that it is observational. The models used show relations after adjusting for available characteristics, but they do not identify causal effects. I cannot observe gender discrimination or institutional decisions related to men's and women's careers, and I do not have the data about parenthood, caring obligations, or career breaks. I am not able to test whether they are factors related to the observed gender differences.

I also do not have a persistent author identifier. I link individuals between years using a proxy ID based on last name and *firstyr*. The results are practically the same when I use alternative author-matching procedures, which is an important test. This does not mean that all individuals were correctly linked, however. It is still possible to have both *false merges* and *false splits*, including the cases when last names were changed.

A similar limitation is related to gender classification. At the basic probability threshold, genderize.io classified about three-quarters of scientists. The percentages of unclassified cases differed across countries, fields, entry cohorts, and persistence levels. Interestingly, the main results were very similar under higher classification thresholds. A manual validation of 300 scientists showed asymmetric errors, with lower precision for female classifications.

Bibliometric data show who is in the top 2%, where they are located within the elite, and whether they return to subsequent annual lists. However, they cannot show the entirety of scientific careers. A longer panel (comprising more years) would provide more data. Also additional data about collaboration patterns, authorship positions in publications, access to financing, institutions and their types and mobility patterns would make it possible to examine in much more detail the origins of the gender differences I observe. The data about family-related issues and career breaks would be especially useful.

The implication is that the threshold alone does not suffice to describe the elite. The threshold allows us to examine who is above it (inclusion), without showing where scientists are located (position) and how long they stay above it (persistence). My data show that women's representation in top 2% is on the rise while the differences in position and persistence remain visible. Representation emerges from this study as only a part of the analysis of gender inequalities in global science.

## 6. Conclusions

In 2019–2024, the representation of women in the global top 2% was systematically rising. Women more often than men appeared as the first-observed in the lists. Once they are within the top 2%, however, women are less likely to be in the top segments of the c-score distribution (position). They are also less likely to be among those scientists who are in the annual lists for several years (persistence).

The gender difference in persistence in elite membership does not disappear after adjustment for academic-entry cohort, country, and broad field. In the fully adjusted model, women have smaller chances of membership in the durable core, which represents appearance in five or six annual lists (OR = 0.886, 95% CI: 0.868–0.905; AME = −2.08 pp). The results differ only slightly when alternative definitions of persistence and other name-matching procedures between years are used.

In sum, knowing who crossed the top 2% threshold is not enough to explain the gender gap in the global citation elite. It is equally important to know how the individuals are located above the threshold and how long they are able to stay there. My data show that the answers are not the same for men and women. Women's representation in the top 2% is rising, but gender differences in position and persistence remain. Therefore, I suggest analyzing inclusion, position, and persistence separately. Elite membership alone, without further specifications, shows only some inequalities at the top of science and hides others.

## Acknowledgments

[…]

## Data Availability

I used data publicly available from Elsevier Data Repository 2026, namely the updated science-wide author databases of standardized citation indicators version 8 (https://elsevier.digitalcommonsdata.com/datasets/btchxktzyw/8).

## Competing Interests

The author declares no competing interests.

## Ethical Approval

This article does not report any studies with human participants performed by the authors.

## Informed Consent

This article does not contain any studies with human participants performed by the author.

# Electronic Supplementary Material to:

# Entering the Elite, But Remaining Outside Its Core?
## *The Temporal Structure of Gender Inequality in the Global Citation Elite*

**Marek Kwiek**
Center for Public Policy Studies, Adam Mickiewicz University, Poznan, Poland
kwiekm@amu.edu.pl, ORCID: orcid.org/0000-0001-7953-1063

**Table S1.** Diagnostics and sensitivity analysis of longitudinal author matching, 2019–2024

**Panel A.** Construction of the annual panel and within-year duplication

| Year | Source records | Unique proxy IDs within year | Duplicate records | Duplicate rate (%) |
|---|---|---|---|---|
| 2019 | 161,441 | 161,132 | 309 | 0.192 |
| 2020 | 190,063 | 189,500 | 563 | 0.296 |
| 2021 | 200,196 | 199,440 | 756 | 0.378 |
| 2022 | 210,198 | 209,244 | 954 | 0.454 |
| 2023 | 223,152 | 221,915 | 1,237 | 0.554 |
| 2024 | 236,313 | 234,858 | 1,455 | 0.616 |
| **Total** | **1,221,363** | **1,216,089**[1] | **5,274** | **0.432** |

[1] The sum of unique scientist-year observations in six annual files; the same scientist can occur in several years.

**Panel B. Panel-level indicators of directly observable matching ambiguity**

| Diagnostic | N | Percentage (%) |
|---|---|---|
| Unique baseline proxy scientists | 465,707 | 100.000 |
| Proxy IDs with ≥1 directly observed within-year collision | 2,223 | 0.477 |
| Proxy IDs without an observed within-year collision | 463,484 | 99.523 |
| Missing authfull or firstyr | 0 | 0.000 |

**Panel C. Geographic concentration of collision-prone proxy IDs**

| Country | Collision-prone IDs | Share of all collision-prone IDs (%) | Proxy scientists in country | Within-country collision rate (%) |
|---|---|---|---|---|
| China | 1,542 | 69.37 | 47,915 | 3.218 |
| The USA | 310 | 13.95 | 149,406 | 0.207 |
| The UK | 68 | 3.06 | 40,178 | 0.169 |
| Australia | 49 | 2.20 | 17,421 | 0.281 |
| India | 44 | 1.98 | 11,928 | 0.369 |
| Hong Kong | 41 | 1.84 | 3,099 | 1.323 |
| Singapore | 27 | 1.21 | 2,487 | 1.086 |
| Canada | 21 | 0.94 | 17,557 | 0.120 |
| Germany | 15 | 0.67 | 20,849 | 0.072 |
| Japan | 13 | 0.58 | 10,180 | 0.128 |
| Pakistan | 12 | 0.54 | 1,812 | 0.662 |
| S. Korea | 11 | 0.49 | 7,818 | 0.141 |
| Sweden | 9 | 0.40 | 5,484 | 0.164 |
| Switzerland | 8 | 0.36 | 7,063 | 0.113 |
| Netherlands | 5 | 0.22 | 10,244 | 0.049 |

| Other | 58 | 2.61 | — | — |
|---|---|---|---|---|
| **Total** | **2,223** | **100.00** | **465,707** | **0.477[2]** |

[2] The collision indicator among 465,707 scientists identified using the baseline proxy ID.

**Panel D. Gender persistence gap under alternative author-matching specifications**

| Matching specification | Known-gender N | Women in durable core (%) | Female durable-core OR (95% CI) |
|---|---|---|---|
| Baseline ID | 347,193 | 18.62 | 0.841 (0.824–0.858) |
| Baseline, collision-prone IDs excluded | 346,364 | 18.68 | 0.844 (0.827–0.861) |
| Normalized ID | 340,181 | 18.66 | 0.843 (0.826–0.859) |
| Normalized ID, collision-prone IDs excluded | 339,358 | 18.72 | 0.845 (0.829–0.862) |

**Source:** Author's calculations based on the *Updated Science-Wide Author Databases of Standardized Citation Indicators*, annual single-year Top 2% datasets for 2019–2024, derived from Scopus data.

The baseline proxy IDs link *authfull* with *firstyr*. Panel A shows deduplication within particular years; duplicates are source records surpassing the number of unique proxy IDs. Panel B shows directly observable collisions, that is, the cases in which the same proxy ID occurs in more than one source record in a given year. Collisions indicate possible ambiguity in name-matching but they are not a full estimation of the author disambiguation error. Panel C ascribes scientists to the most often observable country of affiliation )the modal country) in the years 2019-2024; in the case of equal numbers of affiliations, we use the most recent one. Collision indicators were calculated within the same national populations. Panel D compares four author name-matching procedures and their role in gender gap in durable-core membership. Known gender means their classification with the genderize.io threshold of ≥0,90; durable core means the presence in 5 or 6 annual top 2% lists. OR for women are unadjusted and compare women with men.

**Table S2.** Robustness of the gender persistence gap to alternative author-matching specifications

| Author-matching specification | Known-gender N | Women, N | Women among known gender (%) | Women among one-year members (%) | Women among durable core (%) | Difference (pp) | Female durable-core OR (95% CI) |
|---|---|---|---|---|---|---|---|
| **Baseline ID** | **347,193** | **71,992** | **20.74** | **21.58** | **18.62** | **−2.96** | **0.841 (0.824–0.858)** |
| Baseline, collision-prone IDs excluded | 346,364 | 71,919 | 20.76 | 21.58 | 18.68 | −2.90 | **0.844 (0.827–0.861)** |
| **Normalized ID** | 340,181 | 70,479 | 20.72 | 21.57 | 18.66 | −2.92 | **0.843 (0.826–0.859)** |
| Normalized ID, collision-prone IDs excluded | 339,358 | 70,407 | 20.75 | 21.58 | 18.72 | −2.86 | **0.845 (0.829–0.862)** |

**Source:** Author's calculations based on the *Updated Science-Wide Author Databases of Standardized Citation Indicators*, annual single-year Top 2% datasets for 2019–2024, derived from Scopus data, linked to Genderize.io classifications.

The baseline proxy ID links *authfull* with *firstyr*. Collision-prone IDs are identifiers for which at least a single collision was observed within a year. A normalized identifier is additionally adjusted by Unicode, diacritic marks, punctuation and spaces in linking records between years. One-year members are scientists who are present in exactly one annual list of the Top 2%; durable core includes scientists present in five or six annual lists. We classify gender using genderize.io with the probability threshold of ≥0.90. Difference (pp) means the difference between the share of women in durable core and their share among one-year members. OR for women are adjusted and they compare women with men. OR = odds ratio; CI = 95% confidence interval; pp = percentage points.

The change in the name-matching procedure practically does not change the results. With the baseline proxy ID, women were 21.58% of one-year members and 18.62% of durable core. The difference was −2.96 pp, with OR = 0.841 (95% CI: 0.824–0.858).

After the removal of collision-prone IDs, OR was 0.844. With the normalized identifier, it is 0.843, and after the combination of normalization with the removal of collision cases, it is 0.845. In all four variants, the share of women in durable core remains in a fairly narrow interval of 18.62–18.72%. Thus the way of matching authors has a marginal influence on the result of persistence of being in the citation elite.

**Table S3.** Robustness of the main longitudinal results to alternative genderize.io probability thresholds

| Genderize.io threshold | Classified scientists, N | Coverage (%) | Women (%) | First-observed analytic N | First-observed OR (95% CI) | First-observed AME (pp) | Durable-core analytic N | Durable-core OR (95% CI) | Durable-core AME (pp) | Annual-exit analytic N | Full annual-exit OR (95% CI) |
|---|---|---|---|---|---|---|---|---|---|---|---|
| ≥0.90 | 347,193 | 74.55 | 20.74 | 822,225 | 1.178 (1.164–1.193) | +3.19 | 347,168 | 0.886 (0.868–0.905) | −2.08 | 758,659 | 0.969 (0.956–0.983) |
| ≥0.95 | 328,992 | 70.64 | 20.58 | 779,012 | 1.186 (1.172–1.201) | +3.32 | 328,968 | 0.880 (0.861–0.899) | −2.20 | 720,563 | 0.969 (0.955–0.983) |
| ≥0.99 | 288,360 | 61.92 | 20.94 | 684,852 | 1.203 (1.187–1.219) | +3.55 | 288,338 | 0.873 (0.853–0.893) | −2.35 | 637,732 | 0.968 (0.953–0.983) |

**Source.** Author's calculations based on the *Updated Science-Wide Author Databases of Standardized Citation Indicators*, annual single-year Top 2% datasets for 2019–2024, derived from Scopus data and linked to genderize.io classifications.

We classify gender as men or women when the probability of genderize.io reaches the necessary threshold; we classify all other cases as unknown. Coverage means the share of scientists with binary classification in the total population of 465,707 individuals.

Women (%) refers to the classified population. Analytical N depends on the model: for first-observed presence, these are scientist-year observations, for durable-core membership, these are unique scientists and for annual exit, these are qualified transitions between subsequent years. The model of durable core includes academic-entry cohort, country group and broad field. The model of annual exit additionally includes current c-score(t), prior elite history, and year fixed effects; standard errors are clustered at the level of the scientist. The reference category are men. OR = odds ratio; AME = average marginal effects for women minus for men; CI = 95% confidence interval; pp = percentage points.

Increasing the genderize.io threshold clearly decreases the coverage: from 74.55% with ≥0.90 to 70.64% with ≥0,95 and 61.92% with ≥0.99. The share of women is practically stable. It is 20.74%, 20.58% and 20.94%, respectively.

Also longitudinal results do not change. For first-observed presence, OR increases with the thresholds from 1.178 to 1.186 to 1.203. For durable-core membership, it is 0.886, 0.880 and 0.873. The result for annual exit is almost identical for all three thresholds: 0.969, 0.969 and 0.968.

**Table S4.** Selectivity of unknown-gender classification in the global citation-elite population

**Panel A.** By year

| Year | Total N | Unknown N | Unknown (%) |
|---|---|---|---|
| 2019 | 161,132 | 37,411 | 23.22 |
| 2020 | 189,500 | 42,714 | 22.54 |
| 2021 | 199,440 | 44,543 | 22.33 |
| 2022 | 209,244 | 47,086 | 22.50 |
| 2023 | 221,915 | 50,793 | 22.89 |
| 2024 | 234,858 | 47,596 | 20.27 |

**Panel B.** By country

| Country | Total N | Known gender N | Unknown N | Unknown (%) |
|---|---|---|---|---|
| United States | 149,431 | 122,333 | 27,098 | 18.13 |
| China | 48,329 | 17,238 | 31,091 | 64.33 |
| United Kingdom | 40,156 | 32,391 | 7,765 | 19.34 |
| Germany | 20,846 | 17,445 | 3,401 | 16.31 |
| Canada | 17,544 | 14,055 | 3,489 | 19.89 |
| Australia | 17,396 | 13,924 | 3,472 | 19.96 |
| Italy | 13,975 | 11,057 | 2,918 | 20.88 |
| France | 12,569 | 9,307 | 3,262 | 25.95 |
| India | 11,968 | 8,237 | 3,731 | 31.17 |
| Netherlands | 10,251 | 7,980 | 2,271 | 22.15 |
| Japan | 10,152 | 8,579 | 1,573 | 15.49 |
| Spain | 9,670 | 7,769 | 1,901 | 19.66 |
| South Korea | 7,778 | 3,772 | 4,006 | 51.50 |
| Switzerland | 7,081 | 5,845 | 1,236 | 17.46 |
| Sweden | 5,449 | 4,544 | 905 | 16.61 |
| Iran | 4,369 | 3,697 | 672 | 15.38 |
| Denmark | 4,351 | 3,706 | 645 | 14.82 |
| Belgium | 4,251 | 3,398 | 853 | 20.07 |
| Saudi Arabia | 3,685 | 3,020 | 665 | 18.05 |
| Brazil | 3,662 | 3,129 | 533 | 14.55 |

**Panel C.** By broad field

| Broad field | Total N | Unknown N | Unknown (%) |
|---|---|---|---|
| Enabling & Strategic Technologies | 42,177 | 17,541 | 41.59 |
| Engineering | 35,394 | 13,160 | 37.18 |
| Physics & Astronomy | 42,099 | 15,429 | 36.65 |
| Chemistry | 29,231 | 9,124 | 31.21 |
| Information & Communication Technologies | 29,320 | 8,813 | 30.06 |
| Earth & Environmental Sciences | 16,832 | 5,013 | 29.78 |
| Mathematics & Statistics | 5,579 | 1,621 | 29.06 |
| Agriculture, Fisheries & Forestry | 16,506 | 4,762 | 28.85 |
| Built Environment & Design | 2,594 | 747 | 28.80 |
| Biology | 17,729 | 3,428 | 19.34 |
| Clinical Medicine | 149,612 | 27,364 | 18.29 |
| Biomedical Research | 33,428 | 5,800 | 17.35 |
| Economics & Business | 11,450 | 1,685 | 14.72 |
| Public Health & Health Services | 8,904 | 1,141 | 12.81 |
| Psychology & Cognitive Sciences | 7,896 | 935 | 11.84 |
| Social Sciences | 11,058 | 1,304 | 11.79 |
| Communication & Textual Studies | 2,200 | 258 | 11.73 |
| Historical Studies | 2,258 | 238 | 10.54 |
| Philosophy & Theology | 1,086 | 86 | 7.92 |
| Visual & Performing Arts | 282 | 18 | 6.38 |

**Panel D.** By academic-entry cohort

| Academic-entry cohort | Total N | Unknown N | Unknown (%) |
|---|---|---|---|
| pre-1990 | 153,181 | 32,338 | **21.11** |
| 1990–1999 | 120,503 | 28,951 | **24.03** |
| 2000–2009 | 128,050 | 36,104 | **28.20** |
| 2010+ | 63,973 | 21,121 | **33.02** |
| **Total** | **465,707** | **118,514** | **25.45** |

**Panel E.** By persistence

| Annual appearances | Total N | Known gender N | Unknown N | Unknown (%) |
|---|---|---|---|---|
| 1 year | 202,042 | 141,344 | 60,698 | **30.04** |
| 2 years | 70,432 | 51,816 | 18,616 | 26.43 |
| 3 years | 49,647 | 38,049 | 11,598 | 23.36 |
| 4 years | 43,097 | 34,025 | 9,072 | 21.05 |
| 5 years | 51,080 | 41,031 | 10,049 | 19.67 |
| 6 years | 49,409 | 40,928 | 8,481 | **17.16** |
| **Total** | **465,707** | **347,193** | **118,514** | **25.45** |

**Source:** Author's calculations based on the *Updated Science-Wide Author Databases of Standardized Citation Indicators*, annual single-year Top 2% datasets for 2019–2024, derived from Scopus data, linked to Genderize.io classifications.

Unknown gender includes scientists who under the genderize.io baseline probability threshold ≥0.90 were not classified as men or women. Panel A uses science-year observations and therefore annual frequencies should not be summed in order to have the number of unique scientists. Panels B-E refer to the level of the individual. Country and broad field are determined on the basis of the most often occurring category (the modal value) in annual observations; in the case of equal numbers, we use the most recent observation. Academic-entry cohort is determined on the basis of *firstyr*. Persistence means the number of annual lists of top 2% in which the scientists appeared in the years 2019-2024.

With the genederize.io threshold ≥0.90, 118,514 out of 465,707 scientists (25.45%) are in the unknown category. Differences are large across the countries. In China, the percentage of unknown cases is 64.33%, and in South Korea it is 51.50%. But in the USA, it is 18.13%, and in the UK 19.34%. The percentage exceeds 30% also in several fields, for instance in Engineering, Physics & Astronomy, Chemistry, Information & Communication Technologies and Enabling & Strategic Technologies.

There are also difference between academic-entry cohorts. The share of unknown cases increases from 21.11% in the pre-1990 cohort to 33.02% in the 2010+ cohort.

The most powerful pattern applies to persistence. For scientists present only in a single annual list, the unknown cases are 30.04%. Among scientists who are present in all six annual lists, it is 17.16%, with a decrease at every subsequent level. So the population with the recognized gender is more persistent in presence in the Top 2% than in the full population. This result needs to be taken into account while interpreting gender results.

**Table S5.** Blinded manual validation of genderize.io gender classification.

**Panel A.** Validation sample by genderize.io probability band.

| Probability band | N | Resolved | Unresolved | Correct | Accuracy (%) | 95% CI |
|---|---|---|---|---|---|---|
| ≥0.99 | 123 | 108 | 15 | 107 | 99.07 | 94.94–99.84 |
| 0.95–0.9899 | 58 | 48 | 10 | 41 | 85.42 | 72.83–92.75 |
| 0.90–0.9499 | 61 | 51 | 10 | 39 | 76.47 | 63.24–86.00 |
| <0.90 | 58 | 52 | 6 | 41 | 78.85 | 65.97–87.76 |
| Total | 300 | 259 | 41 | 228 | 88.03 | 83.51–91.44 |

*Note.* N = sample size; Resolved = manually resolved cases; Unresolved = cases without a manual gender classification; Correct = cases in which the Genderize.io and manual classifications agree.

**Panel B.** Classification performance by gender.

| Genderize.io | Resolved | TP | FP | Precision (%) | Recall (%) | F1 |
|---|---|---|---|---|---|---|
| Female | 130 | 102 | 28 | 78.46 | 97.14 | 0.868 |
| Male | 129 | 126 | 3 | 97.67 | 81.82 | 0.890 |
| Overall | 259 | — | — | — | — | — |

*Note.* Resolved = manually resolved cases within each Genderize.io classification category; TP = true positives; FP = false positives. Each gender is treated as the positive category in turn. The counts of 130 and 129 refer to Genderize.io classifications, not manual classifications.

**Panel C.** Validation performance by country/name context.

| Country/context | N | Resolved | Unresolved (%) | Accuracy (%) | 95% CI |
|---|---|---|---|---|---|
| China | 80 | 69 | 13.75 | 76.81 | 65.60–85.19 |
| South Korea | 17 | 16 | 5.88 | 100.00 | 80.64–100.00 |
| Taiwan | 4 | 3 | 25.00 | 66.67 | 20.77–93.85 |
| Hong Kong | 13 | 12 | 7.69 | 91.67 | 64.61–98.51 |
| India | 6 | 5 | 16.67 | 100.00 | 56.55–100.00 |
| Japan | 3 | 1 | 66.67 | 100.00 | 20.65–100.00 |
| United States | 70 | 65 | 7.14 | 90.77 | 81.29–95.70 |
| United Kingdom | 14 | 11 | 21.43 | 90.91 | 62.26–98.38 |
| Western Europe | 21 | 16 | 23.81 | 100.00 | 80.64–100.00 |
| Singapore | 8 | 7 | 12.50 | 85.71 | 48.69–97.43 |
| Malaysia | 4 | 3 | 25.00 | 33.33 | 6.15–79.23 |
| Other countries | 60 | 51 | 15.00 | 94.12 | 84.08–97.98 |
| Total | 300 | 259 | 13.67 | 88.03 | 83.51–91.44 |

*Note.* N = sample size; Resolved = manually resolved cases. Unresolved (%) is calculated among all reviewed cases in each group; accuracy is calculated among resolved cases. CI = confidence interval.

**Panel D.** Confusion matrix for manually resolved cases.

| | Manual: Female | Manual: Male | Total |
|---|---|---|---|
| Genderize: Female | 102 | 28 | 130 |
| Genderize: Male | 3 | 126 | 129 |
| **Total** | **105** | **154** | **259** |

For *female* as a positive category: TP = 102, FP = 28, FN = 3, TN = 126.

Manual validation was conducted for the first 300 cases in the randomized order of an original stratified sample of 500 scientists. The remaining 200 cases were not reviewed. The original sample was stratified by genderize.io probability band, predicted gender, and country/name context, with oversampling of contexts with lower coverage. The human validator's did not know genderize.io predictions or probability scores.

To define gender, the validator used only the textual evidence from reliable institutional or professional websites (photographs and name-based inference were not used). Calculations used the male/ female/ unresolved labels. All 259 resolved cases (male/female) are accompanied by URLs. The 41 unresolved cases are reported separately and excluded from accuracy, precision, recall, and F1 calculations.

The four panels show unweighted results. Accuracy means the number of matching genderize.io and manual labels divided by the number of manually resolved cases. Manually resolved N is counted within each genderize.io prediction category. In Panel C, Western Europe includes the sampled cases from Austria, France, Germany, the Netherlands, and Switzerland. Singapore, Malaysia, and Other countries are the remaining cases not covered by the other named categories. The categories are mutually exclusive.

**Table S6.** Robustness of the gender gap using alternative definitions of persistent elite membership

| Definition of persistence | Women: observed membership (%) | Men: observed membership (%) | Adjusted OR for women (95% CI) | Adjusted P: Women (%) | Adjusted P: Men (%) | AME (pp) | 95% CI for AME |
|---|---|---|---|---|---|---|---|
| Primary durable core: 5–6 annual appearances | 21.20 | 24.24 | 0.886 (0.868–0.905) | 21.95 | 24.03 | −2.08 | −2.43 to −1.72 |
| Consecutive durable core: ≥5 consecutive years | 18.30 | 20.95 | 0.892 (0.872–0.912) | 18.99 | 20.76 | −1.78 | −2.11 to −1.44 |
| Strict six-year core: all 6 annual lists | 10.06 | 12.24 | 0.851 (0.828–0.876) | 10.53 | 12.10 | −1.57 | −1.84 to −1.31 |

**Source**. Author's calculations based on the *Updated Science-Wide Author Databases of Standardized Citation Indicators*, annual single-year Top 2% datasets for 2019–2024, derived from Scopus data, and genderize.io classifications.

The population includes scientists defined as men or women with the genderize.io probability threshold of ≥0.90. Durable core includes scientists present in five or six annual lists of the Top 2% in 2019-2024. Models include academic-entry cohort, country group and broad field.

The reference category in the models is men. Adjusted probabilities are obtained using marginal standardization. The results differ when more restrictive definitions of persistence are used. With the baseline definition, 21.20% of women and 24.24% of men belonged to the durable core. After the adjustment, the results were OR = 0.886 (95% CI: 0.868–0.905), and AME = −2.08 pp.

The requirement of five consecutive years led to OR = 0.892 and AME = −1.78 pp. Under the most restrictive definition – the presence in all six years – the results were as follows: OR = 0.851, and AME

= −1.57 pp. Under each of the three definitions, women have smaller adjusted chances to obtain durable membership in the elite.

**Table S7.** Men and women in the first appearance in the global citation elite, 2020–2024

**Panel A.** Year-specific estimates

| Year | Known-gender N | First-observed N | Men first-observed (%) | Women first-observed (%) | Difference (pp) | OR for women (95% CI) |
|---|---|---|---|---|---|---|
| **2020** | 146,786 | 54,907 | 36.43 | **41.48** | **+5.06** | **1.237 (1.205–1.270)** |
| **2021** | 154,897 | 34,706 | 21.80 | **24.88** | **+3.09** | **1.188 (1.154–1.224)** |
| **2022** | 162,158 | 33,896 | 20.37 | **23.03** | **+2.67** | **1.170 (1.137–1.205)** |
| **2023** | 171,122 | 30,648 | 17.32 | **20.16** | **+2.84** | **1.205 (1.170–1.241)** |
| **2024** | 187,262 | 69,315 | 36.46 | **39.12** | **+2.66** | **1.120 (1.094–1.146)** |

**Panel B.** Pooled models with and without 2024

| Observation window | Analytic N | First-observed N | Adjusted P: Men (%) | Adjusted P: Women (%) | AME (pp) | OR for women (95% CI) |
|---|---|---|---|---|---|---|
| **2020–2024** | **822,225** | **223,472** | **26.54** | **29.73** | **+3.19** | **1.178 (1.164–1.193)** |
| **2020–2023** | **634,963** | **154,157** | **23.61** | **26.97** | **+3.36** | **1.202 (1.185–1.219)** |

**Source:** Author's calculations based on the *Updated Science-Wide Author Databases of Standardized Citation Indicators*, annual single-year Top 2% datasets for 2019–2024, derived from Scopus data, linked longitudinally at the scientist level and to Genderize.io classifications.

In each year from 2020 to 2024, women had higher probability of a first observed appearance in our panel than man. Annual differences were between +2.66 and +5.06 pp, and OR differed in the range of 1.120-1.237.

**Table S8.** Sequential logistic regression models of annual exit from the global citation elite, 2019–2024

**Panel A.** Gender coefficient across sequential model specifications

| Model | Specification | Analytic N | Women OR (95% CI) | Clustered SE | *p* | AME (pp) | 95% CI for AME |
|---|---|---|---|---|---|---|---|
| **M1** | Gender + year FE | 758,684 | 1.064 (1.049–1.080) | 0.00737 | <.001 | +1.28 | +0.98 to +1.58 |
| **M2** | M1 + current *c-score(t)* | 758,684 | 1.022 (1.008–1.037) | 0.00736 | .003 | +0.44 | +0.15 to +0.73 |
| **M3** | M2 + prior elite history | 758,684 | 0.990 (0.977–1.004) | 0.00690 | .158 | −0.18 | −0.44 to +0.07 |
| **M4** | M3 + academic-entry cohort + country group + broad field | 758,659 | 0.969 (0.956–0.983) | 0.00718 | <.001 | −0.58 | −0.84 to −0.32 |

**Panel B.** Gender × current citation position interaction

| Model | Parameter | OR (95% CI) | Clustered SE | *p* |
|---|---|---|---|---|
| **M5** | Woman | 0.971 (0.958–0.985) | 0.00724 | <.001 |
| | *c-score(t)*, per +1 SD | 0.750 (0.745–0.755) | 0.00355 | <.001 |
| | Woman × *c-score(t)* | 1.016 (1.001–1.032) | 0.00772 | .038 |

**Panel C.** Model structure

| **Covariate** | **M1** | **M2** | **M3** | **M4** | **M5** |
|---|---|---|---|---|---|
| Woman | ✓ | ✓ | ✓ | ✓ | ✓ |
| Year fixed effects | ✓ | ✓ | ✓ | ✓ | ✓ |
| Year-standardized *c-score(t)* | — | ✓ | ✓ | ✓ | ✓ |
| Prior elite history | — | — | ✓ | ✓ | ✓ |
| Academic-entry cohort | — | — | — | ✓ | ✓ |
| Country group | — | — | — | ✓ | ✓ |
| Broad field | — | — | — | ✓ | ✓ |
| Woman × *c-score(t)* | — | — | — | — | ✓ |
| Scientist-clustered SE | ✓ | ✓ | ✓ | ✓ | ✓ |

Additional coefficients relevant to sequential attenuation

| **Model** | **Predictor** | **OR (95% CI)** | **Clustered SE** | ***p*** |
|---|---|---|---|---|
| **M2** | *c-score(t)*, per +1 SD | **0.701 (0.697–0.705)** | **0.00323** | **<.001** |
| **M3** | *c-score(t)*, per +1 SD | **0.810 (0.805–0.815)** | **0.00302** | **<.001** |
| **M3** | Prior elite history, per additional prior year | **0.617 (0.614–0.620)** | **0.00286** | **<.001** |
| **M4** | *c-score(t)*, per +1 SD | **0.752 (0.747–0.757)** | **0.00327** | **<.001** |
| **M4** | Prior elite history, per additional prior year | **0.592 (0.588–0.595)** | **0.00294** | **<.001** |

**Source:** Author's calculations based on the *Updated Science-Wide Author Databases of Standardized Citation Indicators*, annual single-year Top 2% datasets for 2019–2024, derived from Scopus data, linked longitudinally at the scientist level and to Genderize.io classifications.

Annual exit means the presence in the Top 2% list in the year t and the absence in the list in the year t+1. We observe 5 transitions, from 2019→2020 to 2023→2024. M1 includes gender year fixed effects. M2 Additionally includes the current c-score(t), adjusted within each year; OR for this variable refers to a 1-standard deviation increase. M3 adds prior elite history measured by the number of prior annual lists in which the scientist appeared. M4 adds academic-entry cohort, country group and broad field, and M5 adds the Woman × c-score(t) interaction. Analytical N are qualifiable transitions with complete data or variables included in a given model. Standard errors are clustered at the level of the scientist. The reference category is men. OR = odds ratio; AME = average marginal effect for women minus for men; CI = 95% confidence interval; SE = standard error; FE = fixed effects; pp = percentage points.

**Table S9.** Gender composition across within-elite *c-score* position in the global citation elite, 2024

**Panel A.** Women's representation across within-elite *c-score* deciles

| **Within-elite *c-score* decile** | **Women, N** | **Men, N** | **Known-gender N** | **Women (%)** | **Difference from D1 (pp)** |
|---|---|---|---|---|---|
| D1 — lowest | 5,034 | 13,693 | 18,727 | **26.88** | Reference |
| D2 | 4,073 | 14,653 | 18,726 | **21.75** | −5.13 |
| D3 | 3,876 | 14,850 | 18,726 | **20.70** | −6.18 |
| D4 | 4,090 | 14,636 | 18,726 | **21.84** | −5.04 |
| D5 | 3,996 | 14,730 | 18,726 | **21.34** | −5.54 |
| D6 | 4,035 | 14,691 | 18,726 | **21.55** | −5.33 |
| D7 | 3,966 | 14,760 | 18,726 | **21.18** | −5.70 |
| D8 | 3,743 | 14,983 | 18,726 | **19.99** | −6.89 |
| D9 | 3,376 | 15,350 | 18,726 | **18.03** | −8.85 |
| D10 — highest | 2,852 | 15,875 | 18,727 | **15.23** | **−11.65** |
| **Total** | **39,041** | **148,221** | **187,262** | **20.85** | — |

**Panel B.** Continuous model of within-elite *c-score* position

| Predictor | Analytic N | Scaling | OR for women | 95% CI | *p*-value |
|---|---|---|---|---|---|
| Within-elite *c-score* percentile | **187,262** | +10 percentile points | **0.951** | **0.947–0.954** | **<.001** |

**Source:** Author's calculations based on the *Updated Science-Wide Author Databases of Standardized Citation Indicators*, 2024 single-year Top 2% dataset, derived from Scopus data and linked to genderize.io classifications.

Panel A divides the distribution of c-score within the elite in 2024 into deciles, from D1 (the lowest) to D10 (the highest). The deciles have approximately the same frequency. Women (%) means the share of women among scientists with recognized gender; for difference from D1 (pp), the reference category is D1.

Panel B uses c-score percentile as a continuous variable. OR refers to the move by 10 percentile points upwards within the internal c-score distribution. The reference category are men. OR = odds ratio; CI = 95% confidence interval; pp = percentage points.

Women are 26.88% scientists in D1 and 15.23% in D10, which means the difference of −11.65 pp. The decline is not monotonic. Between D2 and D8, the share of women is generally at the level of approximately 20-22% and then it drops to 18.03% in D9 and 15.23% in D10.

The continuous model shows the same general direction. The change of 10 percentile points upwards and the distribution of c-score corresponds to OR = 0.951 (95% CI: 0.947–0.954; $p < 0.001$; N = 187,262) for women.

**Table S10.** Gender representation, membership overlap, and cohort composition in annual and career-long citation elites, 2019–2024

**Panel A.** Gender representation in annual and career-long citation elites, 2019–2024

| Year | Annual t | Annual k gender N | Annual v N | Annual v (%) | Career-lc total N | Career-lc known-g | Career-lc women N | Career-lc women (' | Career-lc annual (p |
|---|---|---|---|---|---|---|---|---|---|
| 2019 | 161,132 | 123,721 | 22,698 | **18.35** | 159,577 | 91,710 | 13,987 | **15.25** | **−3.10** |
| 2020 | 189,500 | 146,786 | 28,404 | **19.35** | 186,046 | 105,731 | 17,126 | **16.20** | **−3.15** |
| 2021 | 199,440 | 154,897 | 30,453 | **19.66** | 194,817 | 110,920 | 18,019 | **16.25** | **−3.41** |
| 2022 | 209,244 | 162,158 | 32,671 | **20.15** | 204,454 | 118,123 | 19,587 | **16.58** | **−3.56** |
| 2023 | 221,915 | 171,122 | 35,393 | **20.68** | 216,858 | 125,090 | 21,284 | **17.01** | **−3.67** |
| 2024 | 234,858 | 187,262 | 39,041 | **20.85** | 230,062 | 133,846 | 22,868 | **17.08** | **−3.76** |

**Panel B.** Membership overlap between annual and career-long citation elites, 2024

| Membership status | Women, N | Men, N | Known-gender N | Women (%) | Share of union (% |
|---|---|---|---|---|---|
| Annual only | **18,602** | **50,519** | **69,121** | **26.91** | **34.06** |
| Career-long only | **2,429** | **13,276** | **15,705** | **15.47** | **7.74** |
| Both annual and career-lo | **20,439** | **97,702** | **118,141** | **17.30** | **58.21** |
| **Union** | **41,470** | **161,497** | **202,967** | **20.43** | **100.00** |

The membership overlap can additionally be summarized using the Jaccard index:

$$J(A,C)=\frac{|\,A\cap C\,|}{|\,A\cup C\,|}=\frac{118{,}141}{202{,}967}=0.5821.$$

Thus, 58.21% of the known-gender union belonged to both the annual and career-long citation elites.

**Panel C.** Observed and cohort-standardized women's representation, 2024

| Elite definition | Analytic N | Observed women (%) | Standardized women (%) | Effect (pp) |
|---|---|---|---|---|
| Annual elite | 187,262 | 20.85 | 20.63 | −0.21 |
| Career-long elite | 133,846 | 17.09 | 17.38 | +0.30 |
| **Career-long − annual** | — | **−3.76** | **−3.25** | **+0.51** |

**Source**. Author's calculations based on the *Updated Science-Wide Author Databases of Standardized Citation Indicators*, annual single-year and career-long Top 2% datasets for 2019–2024, derived from Scopus data and linked to Genderize.io classifications.

The annual global citation elite is determined on the basis of citation indicators for a single year while career-long elite is determined on the basis of accumulated indicators from the whole career from the relevant database edition. Gender is classified using genderize.io with the probability threshold of ≥0,90. Panel A composed the share of women in both populations.

Panel B uses the sum of sets of annual and career-long elite from 2024 for scientists with recognized gender. The categories of annual only elite, career-long only elite and both are mutually exclusive. The Jaccard index is the size of the intersection divided by the size of the union of the sample sets

Panel C includes scientists with recognized gender and correct value of *firstyr*. The annual and career-long elite populations are standardized to the same distribution of academic-entry cohort, distinguishing between four cohorts: pre-1990, 1990–1999, 2000–2009 and 2010+. Scientists belonging to both elites enter the reference population only once. Standardization effect (pp) Means the difference between the standardized and observed share of women. The percentage weakening of the difference is calculated on the basis of and the rounded values as

$$100\times\frac{|\,G_{\text{raw}}\,|-|\,G_{\text{std}}\,|}{|\,G_{\text{raw}}\,|},$$

where $G_{\text{raw}}$i $G_{\text{std}}$ mean differences between annual and career-long elite before and after standardization.

Panel C requires correct firstyr and therefore is analytic population is slightly different from the population in panel A. pp = percentage points.

The share of women was increasing in both populations between 2019 and 2024 but in each year, it was higher in the annual elite (Table S10, Panel A). In the annual elite, the share increased from 18.35% to 20.85%, and in the career-long elite, it increased from 15.25% to 17.08%. The difference between them increased from 3.10 to 3.76 pp.

Both populations are only partially overlapping (Panel B). In 2024, 58.21% of scientists from the sum of the two sets belongs to both elites. 34.06% of scientists belonged only to the annual elite and 7.74% belonged only to the career-long elite. The gender composition of the two groups is clearly different. Women were 26.91% of scientists in the annual only elite, compared with 15.47% in the career-long only group and 17.30% among scientists belonging to both groups.

The cohort structure explains part of this difference (Panel C). Before standardization, women were 20.85% of the annual elite and 17.09% of the career long elite. After adding to both populations the same academic entry-cohort distribution, the shares are 20.63% and 17.38%, respectively. The gender difference decreases from 3.76 to 3.25 pp, that is by 0,51 pp (13.63%).

**Table S11.** Operational definitions of the main analytical constructs

| Construct | Operational definition | Unit of analysis | Observation window / timing | Role in analysis |
|---|---|---|---|---|
| **Annual elite membership** | Presence on the corresponding annual single-year global Top 2% list | Scientist-year | 2019–2024 | Basic measure of inclusion |
| **First-observed presence** | First appearance of a scientist within the observed annual panel | Scientist-year | 2020–2024; 2019 excluded because all members are mechanically first observed | Main outcome for the first-observed analysis |
| **Continuing membership** | Presence in year *t* after at least one earlier observed appearance in the panel | Scientist-year | 2020–2024 | Descriptive complement to first-observed presence |
| **Annual entry** | Presence in year *t* following absence from the immediately preceding annual Top 2% list in year $t{-}1$ | Scientist-year transition | Five transitions ending in 2020–2024 | Descriptive transition measure of annual inflow |
| **Within-elite position** | Percentile position of *c-score* within the relevant annual Top 2% population | Scientist-year | Measured within year | Main measure of position |
| ***c-score* decile** | Ten groups of within-elite *c-score* position; D1 = lowest and D10 = highest | Scientist-year | Within the analyzed annual population | Descriptive analysis of vertical stratification |
| **Academic age** | Observation year minus *firstyr* | Scientist-year | Time-varying | Career-stage measure |
| **Academic-entry cohort** | *firstyr*: pre-1990; 1990–1999; 2000–2009; 2010+ | Scientist | Time-invariant | Control and heterogeneity dimension |
| **Persistence** | Number of annual Top 2% lists on which a scientist appears | Scientist | 2019–2024 | Temporal dimension of elite membership |
| **One-year membership** | Appearance on exactly one of the six annual lists | Scientist | 2019–2024 | Low-persistence reference category |
| **Durable core** | Appearance on five or six of the six annual lists, irrespective of continuity | Scientist | 2019–2024 | Primary persistence outcome |
| **Consecutive durable core** | Longest uninterrupted spell of at least five consecutive annual appearances | Scientist | 2019–2024 | Alternative persistence definition |
| **Strict six-year core** | Appearance on all six annual lists | Scientist | 2019–2024 | Most restrictive persistence definition |
| **Annual exit** | Presence in year *t* followed by absence in year $t{+}1$ | Scientist-year transition | Five transitions, 2019→2020 through 2023→2024 | Transition outcome |
| **Current citation position, *c-score(t)*** | Within-year *c-score* position measured in year *t* and standardized within year for the annual-transition models | Scientist-year transition | Measured before exit outcome at $t{+}1$ | Time-ordered predictor in exit models |
| **Prior elite history** | Number of earlier annual Top 2% lists on which the scientist appeared before year *t* | Scientist-year transition | Measured before the $t \rightarrow t+1$ transition | Time-ordered predictor in exit models |
| **Broad field** | Modal Science-Metrix broad field across annual observations; most recent category resolves ties | Scientist | 2019–2024 | Control and heterogeneity dimension |
| **Country** | Modal country affiliation across annual observations; most recent category resolves ties | Scientist | 2019–2024 | Descriptive contextual variable |
| **Country group** | Fifteen largest national systems in the primary known-gender population; all others combined as *Other* | Scientist | 2019–2024 | Control and heterogeneity dimension |
| **Gender** | Female or male classification from Genderize.io at probability ≥0.90 | Scientist | Person-level | Primary explanatory variable |
| **Unknown gender** | No female/male classification meeting the ≥0.90 probability threshold | Scientist | Person-level | Coverage/selectivity diagnostic |
| **Baseline author ID** | lower(authfull) + firstyr | Scientist | Across annual files | Primary longitudinal linkage |
| **Normalized author ID** | Normalized author-name representation + *firstyr* | Scientist | Across annual files | Author-matching sensitivity analysis |

**Source.** Author's operationalization based on the *Updated Science-Wide Author Databases of Standardized Citation Indicators*, annual single-year Top 2% datasets for 2019–2024, derived from Scopus data, and genderize.io classifications.

First-observed presence means the first appearance in the 2019-2024 panel. Annual entry means something else: the scientist is present in the year $t$ and they are present on the list in the year $t-1$. In 2021 and later, it may also include scientists who appeared earlier in the panel and then returned after a period of absence. Persistence refers to repeated presence in the six observed years. It does not measure elite status in the whole career. Annual exit means the absence in the subsequent Top 2% list. The current c-score(t) and prior elite history are measured before the transition $t \rightarrow t+1$. Country and broad field are the most frequently occurring (modal) classifications at the individual level; in the case of equal numbers, we use the most recent observation. Gender is defined with the genderize.io probability threshold of ≥0.90. Higher probability thresholds are used in sensitivity analyses.